\documentclass[pre,twocolumn,superscriptaddress,amsmath,amssymb,nofootinbib]{revtex4-1}
\usepackage{graphicx,enumitem}
\usepackage{multirow,dcolumn}
\usepackage{txfonts}
\usepackage{hyperref}
\usepackage{soul,cancel}
\usepackage{epsf,amsmath,amssymb,verbatim,color,multirow,pifont,graphicx,cleveref,tabularx}
\usepackage[dvipsnames]{xcolor}
\usepackage{rotating}
\usepackage{amsmath,amsfonts,amssymb,bm,cancel}

\usepackage{tikz}

\begin{document}
\title{Tricritical directed percolation with long-range interaction in one and two dimensions}

\author{Minjae Jo}
\affiliation{CCSS, CTP and Department of Physics and Astronomy, 
Seoul National University, Seoul 08826, Korea}

\author{B. Kahng}
\email{bkahng@snu.ac.kr}
\affiliation{CCSS, CTP and Department of Physics and Astronomy, 
Seoul National University, Seoul 08826, Korea}


\begin{abstract}
Recently, the quantum contact process, in which branching and coagulation processes occur both coherently and incoherently, was theoretically and experimentally investigated in driven open quantum spin systems. In the semi-classical approach, the quantum coherence effect was regarded as a process in which two consecutive atoms are involved in the excitation of a neighboring atom from the inactive (ground) state to the active state (excited $s$-state). In this case, both second-order and first-order transitions occur. Therefore, a tricritical point exists at which the transition belongs to the tricritical directed percolation (TDP) class. On the other hand, when an atom is excited to the $d$-state, long-range interaction is induced. Here, to account for this long-range interaction, we extend the TDP model to one with long-range interaction in the form of $\sim 1/r^{d+\sigma}$ (denoted as LTDP), where $r$ is the separation, $d$ is the spatial dimension, and $\sigma$ is a control parameter. In particular, we investigate the properties of the LTDP class below the upper critical dimension $d_c=$ min$(3,\,1.5\sigma)$. We numerically obtain a set of critical exponents in the LTDP class and determine the interval of $\sigma$ for the LTDP class. Finally, we construct a diagram of universality classes in the space ($d$, $\sigma$).
\end{abstract}



\maketitle


\section{Introduction}
\label{sec:introduction}
In statistical physics, nonequilibrium phase transitions into an absorbing state are a well-known phenomenon and have been widely studied~\cite{marro, grassberger_intro, harris, kinzel, ziff, dickman, obukhov, cardy,hinrichsen, henkel, odor}. One of the most popular models is a contact process (CP). In the CP model, the system contains either an active or an inactive particle at each site of a $d$-dimensional lattice. An active particle activates an inactive particle at the nearest-neighbor site with probability $\kappa$; otherwise, it becomes inactive itself with probability $1-\kappa$. By contrast, an inactive particle cannot recover to an active particle alone. When $\kappa$ is small, inactive particles become more abundant with time, and eventually the system is fully occupied by inactive particles. Then, the system is no longer dynamic and falls into an absorbing state. When $\kappa$ is large, the system remains in an active state with a finite density of active particles. Thus, the CP model exhibits a phase transition from an active to an absorbing state as the control parameter $\kappa$ is decreased in any spatial dimension. This absorbing transition is second-order and belongs to the so-called directed percolation (DP) universality class~\cite{grassberger_intro,obukhov,cardy,hinrichsen,henkel,grassberger_conjecture,janssen_conjecture}. In the DP class, the mean-field solution is valid above the upper critical dimension $d_c=4$. The CP model can be applied to diverse phenomena such as the epidemic spread of infectious disease and the reaction-diffusion process of interacting particles.  

The CP model has been modified in various ways to describe different phenomena. For instance, L\"ubeck introduced the so-called tricritical CP (TCP) model as follows. In addition to the ordinary CP, a pair of consecutive active particles can activate an inactive particle at a nearest-neighbor site with probability $\omega$~\cite{ohtsuki1,ohtsuki2,grassberger,lubeck, windus, windus_1D}. 
The TCP model exhibits an absorbing transition, which is either first-order or second-order depending on the parameters ($\kappa$, $\omega$). The two types of phase boundaries meet at a tricritical point. The absorbing transition at the tricritical point is second-order, and its critical behavior, which is denoted as tricritical DP (TDP), is distinct from that of the DP class. The TDP class has been extensively studied, and various features have been identified. Using the field theoretical approach~\cite{ohtsuki1}, the critical exponents of the TDP class were determined, together with the upper critical dimension, $d_c=3$~\cite{ohtsuki1,ohtsuki2,janssen_TDP}. Moreover, extensive numerical simulations were performed in two dimensions in Refs.~\cite{lubeck, grassberger, windus} using slightly different models. However, the simulations yielded critical exponents that were inconsistent with each other, which was attributed to the inaccuracy of the numerical value of the tricritical point~\cite{odor}. It was also argued that the first-order transition does not occur in the one-dimensional DP-type model~\cite{hinrichsen_first}. Thus, the lower critical dimension seems to be two. 

Recently, the TDP class has attracted considerable attention from the physics community after the quantum contact process (QCP), which belongs to the TDP universality class {in the mean-field semi-classical limit}, was {investigated and realized experimentally in a dissipative quantum system of Rydberg atoms in the presence of the strong dephasing~\cite{gutierrez}.} An active (inactive) particle is represented by a Rydberg atom in an excited state (the ground state). An inactive particle is activated by detuning the excitation energy of an active particle, in a process called antiblockade~\cite{detuning,detuning2,detuning3}. This antiblockade dynamics can be implemented incoherently when strong dephasing noise is applied. In this case, the quantum coherence becomes negligible, and the dynamics is reduced to the classical CP process, which generates a second-order transition. However, when quantum coherence is essential, this case is called the QCP, and it yields second-order and first-order transitions~\cite{marcuzzi,buchhold}. Competition between the two types of processes leads to a tricritical point, which yields another second-order transition that belongs to the TDP {class}. 

\begin{figure}
\includegraphics[width=0.85\columnwidth]{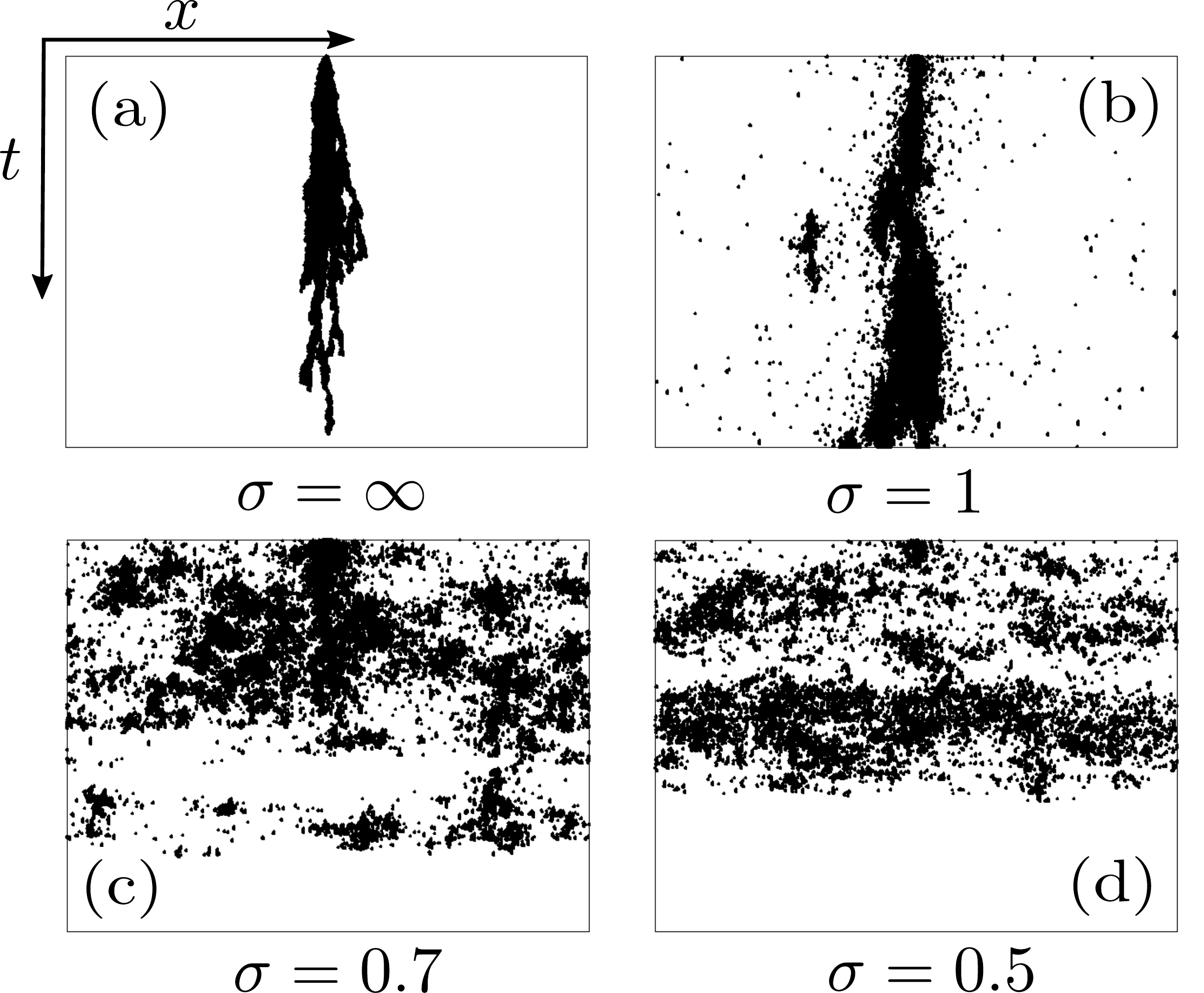}
\caption{Snapshot of active sites of the LTCP model in one dimension at a critical point $(\kappa_c(\omega),\,\omega)$ at a fixed $\omega=0.5 < \omega_c$ (a) and (b) and at the tricritical point ($\kappa_t,\,\omega_t$) (c) and (d). For $\sigma > 1$, the tricritical point does not exist.}
\label{fig:fig1}
\end{figure}

We remark that if an atom is excited to the $s$-state by the QCP, then quantum coherence would occur locally, so the short-range TCP (STCP) model {[Fig.~\ref{fig:fig1}(a)]} would be relevant, which is equivalent to ordinary TCP. On the other hand, if excitation to the $d$-state occurs, dipole--dipole interactions become effective, and a long-range TCP (LTCP) model {[Fig.~\ref{fig:fig1}(b), (c), and (d)]} would be relevant. Although the STCP model has been extensively investigated not only in the mean-field limit but also for low-dimensional cases, the LTCP model has only a mean-field solution~\cite{lastWork}.    

In phase transitions, the interaction range is an essential factor determining the universality class of phase transitions in both equilibrium~\cite{fisher,sak,luijten,longRangeEQ4,longRangeEQ5,longRangeEQ6,longRangeEQ7,longRangeEQ8,horita} and nonequilibrium systems~\cite{LDP1,LDP2, LDP3,LDP4,LDP5,LDP6,linder}. Thus, the classical CP model with long-range interactions was introduced, motivated by the fact that epidemic diseases can be spread by, for instance, L\'evy flight. In this model, the activation process is realized by assigning the probability $\kappa P_I(r)$ that each active particle activates an inactive particle at distance $r$. Thus, $P_I(r)$ represents the probability that a particle at distance $r$ is chosen. $P_I(r)$, which follows the power-law $\sim 1/r^{d+\sigma}$, is non-trivial, where $\sigma>0$ is a control parameter. 

This long-range CP (LCP) exhibits $\sigma$-dependent critical behavior, which is relevant within the interval {denoted as} [$\sigma_{c1}$, $\sigma_{c2}$]. Below $\sigma_{c1}$, the critical behavior is consistent with the mean-field solution. Using the field-theoretical approach, $d_c$ is determined as min$(4,\,2\sigma)${~\cite{LDP1,LDP2}}. Thus, for $\sigma<2$ or $d<4$, $d_c=2\sigma$, and $\sigma_{c1}=d/2$ for $d < 4$. Above $\sigma_{c2}$, it belongs to the ordinary DP class. 
Field-theoretical analysis revealed that $\sigma_{c2}=d+z(1-2\delta)$, where $z$ is a dynamic exponent, and $\delta$ is the critical exponent for the density of active particles $\rho_a(t)\sim t^{-\delta}$ of the ordinary DP class. When $z$ and $\delta$ were replaced with their DP values, $\sigma_{c2}$ was found to be $2.0766$ in one dimension, $2.1725$ in two dimensions, and $2.126$ in three dimensions. However, direct simulation data in one dimension could not reproduce the value $\sigma_{c2}\approx 2.08$, so further investigation is needed in future work to resolve this inconsistency~\cite{LDP1}. 
For $d > 4$, there exists one threshold, $\sigma_c=2$, such that for $\sigma < 2$, the mean-field solution of the long-range DP is valid, whereas for $\sigma > 2$, the mean-field solution of the ordinary DP is valid.  

We focus on the LTCP model. In our previous work, we constructed a phase diagram based on the mean-field solution, which is valid for $d > d_c=$ min$(3,1.5\sigma)${~\cite{lastWork}}. In this case, there exists a characteristic value $\sigma_c${$=2$} such that for $\sigma < \sigma_{c}$, the mean-field solution {of the LTCP} is relevant, and for $\sigma > \sigma_{c}$, the LTCP model behaves like the STCP model. We will show later that when {$d<3$}, the LTCP model exhibits distinctive behavior (characterized as that of the LTCP class) in the interval [$\sigma_{c1}$, $\sigma_{c2}$], where $\sigma_{c1}=2d/3$ because $d_c=${$1.5\sigma$}, and $\sigma_{c2}$ is determined by the hyperscaling relation $\sigma_{c2}=d+z(1-\delta-\delta^\prime)$, where {$\delta^\prime$ is the critical exponent for the survival probability $P(t)\sim t^{-\delta^\prime}$.} We need to replace $z$, $\delta$, and $\delta^\prime$ in the formula with the numerical values of the short-range TDP (STDP) to obtain $\sigma_{c2}$. For $\sigma < \sigma_{c1}$ shown in Fig.~\ref{fig:fig1}(d), the mean-field behavior of the LTDP class appears, and for $\sigma > \sigma_{c2}$ shown in Fig.~\ref{fig:fig1}(a), the behavior of the STDP class appears. The universality class diagram will be shown later. As in the LCP model, the value of $\sigma_{c2}$ is obtained from the hyperscaling relation; however, it is not consistent with the value obtained directly from numerical simulations. Finally, we determine the critical exponents of the LTCP model in the interval [$\sigma_{c1},\,\sigma_{c2}$], which vary continuously with $\sigma$. 

The remainder of this paper is organized as follows. 
In Sec.~\ref{sec:2}, we present the rules of the long-range TCP in detail. 
In Sec.~\ref{sec:3}, the critical behavior of the absorbing transition is determined. 
In Sec.~\ref{sec:4}, we set up the Langevin equation to derive the scaling relation.
In Sec.~\ref{sec:5}, we report numerical results for the long-range TCP. 
In the final section, a summary and discussion are presented.

\section{LTCP model}\label{sec:2}

\begin{table*}[ht]
\caption{Reaction schemes of the CP, TCP, and LTCP. $A$ ($0$) represents the active (inactive) state. TCP$^*$ denotes the TCP model introduced in Ref.~\cite{lubeck}. When the L\'evy exponent $\sigma \to \infty$ in the LTCP, the LTCP model is reduced to the $m$-TCP used in Sec.~\ref{sec:5-2}. The last column indicates the processes explained in Sec.~\ref{sec:2}. The notation $\cdots$ in the LTCP column represents long-range interactions. $P_I(|{\bm r}-{\bm r^\prime|})\sim 1/|{\bm r}-{\bm r^\prime}|^{d+\sigma}$.
}
\begin{center}
\setlength{\tabcolsep}{12pt}
\newlength\lengtha \setlength\lengtha{3mm} 
\newlength\lengthb \setlength\lengthb{8mm}
{\renewcommand{\arraystretch}{1.5}
\begin{tabular}{@{\hspace*{3mm}}c @{\hspace*{7mm}}c @{\hspace*{12mm}}c @{\hspace*{7mm}}c @{\hspace*{12mm}}c @{\hspace*{5mm}}c @{\hspace*{6mm}}c }
\hline
\hline
\multicolumn{2}{l}{\qquad\qquad CP} & \multicolumn{2}{l}{\qquad\qquad TCP$^*$} & \multicolumn{3}{l}{\qquad\qquad\qquad\qquad LTCP}  \\
\hline
Reaction & Probability & Reaction & Probability & Reaction & Probability & Process\\
\hline
\underline{$A$} $\rightarrow 0$ & $1-\kappa$&\underline{$A$} $\rightarrow 0$ & $(1-\omega)(1-\kappa )$&\underline{$A$} $\rightarrow 0$ & $(1-\omega)(1-\kappa )$ & i-a) \\
\underline{$A$}$0 \rightarrow AA$ & $\kappa$&\underline{$A$}$0 \rightarrow A A$ & $(1-\omega)\kappa$&\underline{$A$}$\cdots 0 \rightarrow A \cdots A$ & $(1-\omega)\kappa\,P_I(|\bm{r}-\bm{r}'|)$ & i-b)\\
&&\underline{$A0$} $\rightarrow 0 0$ & $\omega(1-\kappa )$&\underline{$A0$} $\rightarrow 0 0$ & $\omega(1-\kappa )$ & ii-a)\\
&&\underline{$A0$} $\rightarrow AA$  & $\omega\kappa$&\underline{$AA$}$\cdots 0 \rightarrow A A\cdots A$  & $\omega\kappa\,P_I(|\bm{r}-\bm{r}'|)$ & ii-b)\\
&&\underline{$AA$}$0 \rightarrow AAA$  & $\omega$&&&\\
\hline
\hline
\end{tabular}}
\label{tab:tab1}
\end{center}
\end{table*}

We perform numerical simulations by extending the algorithm used in Ref.~\cite{lubeck} for the STCP model to the long-range case. Specifically, the model is set up on a $d$-dimensional lattice composed of $L^d$ sites, where $L$ is the lateral size of the system, and each site is in either the active state (denoted as $A$) or the inactive state (denoted as $0$). We use two different initial configurations: i) one site is active, and the others are all inactive, and ii) all sites are active. Each case will be used for different purposes. We use the periodic boundary condition in the simulations. 
At each time step, the following rules are applied.
\begin{enumerate}[label=(\arabic*)]
\item[i)] An active site is chosen randomly from the list of active sites. Its position is denoted as $\bm{r}_0$. With probability $1-\omega$, a long-range CP is performed as follows:
\begin{enumerate}[label=i-\alph*)]
\item With probability $1-\kappa$, the active site chosen in step i) is inactivated.
\item With probability $\kappa$, a site at a distance $r$ from the position $\bm{r}_0$ is selected with probability $P_I(x)$. If this target site is inactive (0), its state is changed to active ($A$).
\end{enumerate}
\item[ii)] An active site is chosen randomly from the list of active sites. Its position is denoted as $\bm{r}_0$. The state of a nearest-neighbor site is checked with probability $\omega$.   
\begin{enumerate}[label=ii-\alph*)]
\item If the neighbor is inactive, then the active site at $\bm{r}_0$ is inactivated with probability $1-\kappa$.
\item If the neighbor is active, then a third site is selected at a distance $r$ from $\bm{r}_0$ with probability $P_I(r)$. If this target is inactive, it is activated with probability $\kappa$. 
\end{enumerate}
\item[iii)] If the number of active sites is zero, the simulation ends. Otherwise, the time $t$ is advanced by $1/N_a$, where $N_a(t)$ is the total number of active sites in the system at time $t$, and the simulation returns to step i).
\end{enumerate}
In this rule, $P_I(r)$ is given as $\sim 1/r^{d+\sigma}$. This model is controlled by three parameters: i) the L\'evy exponent $\sigma>0$ controlling the long-range interaction, 
ii) the probability $\omega$ of checking the nearest-neighbor site before the reaction, and iii) the probability of the branching process $\kappa$. The reactions are summarized in Table~\ref{tab:tab1}.

\section{Critical behavior of the absorbing transition}
\label{sec:3}

Here we introduce the basic physical quantities used to characterize the critical behavior of the absorbing transition. To proceed, we first consider a system in which a single active site is located at ${\bm r}=0$ at time $t=0$, and the remaining sites are inactive. The LTCP begins in this configuration. We measure the following quantities to characterize the criticality of the LTCP: i) the survival probability $P(t)$ (i.e., the probability that the system has not entered in the absorbing state), ii) the number of active sites $N_a(t)$, and iii) the mean square of the distance from the origin $R^2(t)$. That is, $R^2(t)=(1/N_a(t))\sum_{j=1}^{N_a}{\bm r_j^2}$, where $\bm r_j$ is the position of the $j$-th active site. When sufficiently long-range interactions are considered, the arithmetic average of $R^2(t)\equiv \langle |r(t)|^2\rangle$ may be difficult to obtain numerically~\cite{LDP1, LDP2}. The geometric average $R^2(t)=\exp[\langle \ln|r(t)|^2\rangle]$ may be a suitable alternative. 
Second, one may take as the initial configuration that occupied entirely by active sites. Using this initial configuration, iv) the density $\rho_a(t)$ of active sites at time $t$  is measured. 

At the critical point, these quantities exhibit power-law behavior as follows: 
\begin{align}
P(t)\propto t^{-\delta^\prime}\,,\quad 
N_a(t)\propto t^{\eta}\,,\quad
R^2(t)\propto t^{2/z} \,,\quad
\rho_a(t)\sim t^{-\delta}.
\label{eq:powerlaw0}
\end{align}
The mean density of surviving active sites behaves as $\rho_a(t)P(t)=N_a(t)/R^d(t)$. 
Thus, the exponent $\delta$ is related to the other exponents as $\delta=d/z-\eta-\delta^\prime$.
In particular, at the tricritical point, these exponents are denoted as $\delta^\prime_t$, $\eta_t$, $z_t$, and $\delta_t$. Hereafter, we drop the subscript $t$ indicating the tricritical case for brevity unless it is necessary for clarity.

In the supercritical region, $\kappa > \kappa_c$ for each given $\omega < \omega_t$, and $P(t)$ reaches $P_s$ in the steady state, where $P_s\sim (\kappa-\kappa_c)^{\beta^\prime}$. $\rho_a(t)$ behaves similarly to $\rho_a(t)\to \rho_{a,s}\sim (\kappa-\kappa_c)^{\beta}$. The exponents $\beta^\prime$ and $\beta$ are related to $\delta^\prime=\beta^\prime/\nu_{\|}$ and $\delta=\beta/\nu_{\|}$, where the exponent $\nu_{\|}$ is the mean survival time exponent defined in terms of the mean survival time $\tau \sim (\kappa-\kappa_c)^{-\nu_{\|}}$. 
At the tricritical point, $\beta \ne \beta^\prime$ (equivalently, $\delta \ne \delta^\prime$), whereas in the DP class, they are the same. 

We characterize the critical behavior in finite systems using the finite-size scaling (FSS) theory. In this approach, the critical exponents are determined using the data collapse technique for scaling functions. 
{Data collapse technique is achieved by scaling hypothesis in which the large-scale properties are invariant near the tricritical point $(\kappa_t=\kappa_c(\omega_t),\,\omega_t)$ under the following scale transformations.
\begin{align}
\Delta\kappa&\rightarrow s^{-1}\Delta\kappa\,,\quad
\rho_a\rightarrow s^{-\beta}\rho_a\,,\quad
N_a\rightarrow s^{\nu_{\|}\eta}N_a\,,\quad
P\rightarrow s^{-\beta'}P\,,\quad\nonumber\\
\zeta&\rightarrow s^{\nu_{\bot}}\zeta\,,\quad
\tau\rightarrow s^{\nu_{\|}}\tau\,,\quad
\Delta\omega\rightarrow s^{-\phi}\Delta\omega\,,
\end{align}
where $\Delta \kappa=\kappa-\kappa_t$, $\Delta \omega=\omega-\omega_t$, and $s$ is a scale factor and $\nu_{\bot}$ is the spatial correlation exponent defined in terms of the spatial correlation $\zeta\sim (\kappa-\kappa_c)^{-\nu_{\bot}}$. In addition, $\phi$ is a crossover exponent defined as the ratio of the scaling exponent of $\Delta \kappa$ and $\Delta \omega$.}
For instance, at the tricritical point $(\kappa_t,\,\omega_t)$, the average density $\rho_a(t)$ of active sites behaves as $\rho_a(t,N)=s^{\beta}\rho_a(s^{\nu_{\|}}t,s^{\bar{\nu}_{\bot}}N)\,$, where $\bar{\nu}_{\bot}\equiv d\nu_{\bot}$.

When $s^{\nu_{\|}}t=1$ is chosen, $\rho_a(t)= t^{-{\delta}}f_n(tN^{-\bar{z}})$. Similarly, the other quantities are reduced as 
\begin{align}
P(t)= t^{-\delta^\prime}f_p(tN^{-\bar{z}})\,,\quad 
N_a(t)=t^{\eta}f_{N}(tN^{-\bar{z}})\,,
\label{eq:powerlaw1}
\end{align}
where $\bar{z}=z/d$, $z=\nu_{\|}/{\nu}_{\bot}$, and $f_n$, $f_p$, and $f_N$ are scaling functions.

Near the tricritical point, the number of active sites {and the density of active sites} scales as
\begin{align}
N_a(t,\Delta \kappa,\Delta \omega)&= s^{-\nu_{\|}\eta}N_a(s^{\nu_{\|}}t,s^{-1}\Delta \kappa,s^{-\phi}\Delta \omega)\,,\label{eq:powerlaw2}
\\
\rho_a(t,\Delta \kappa,\Delta \omega)&= s^{\beta}\rho_a(s^{\nu_{\|}}t,s^{-1}\Delta \kappa,s^{-\phi}\Delta \omega)\,.
\label{eq:powerlaw2_2}
\end{align}
At $\Delta \omega=0$, by choosing $s^{\nu_{\|}}t=1$, we can reduce Eq.~\eqref{eq:powerlaw2} to
\begin{align}\label{eq:powerlaw3}
N_a(t)=t^{\eta}f_1(t^{1/\nu_{\|}} \Delta \kappa)\,,
\end{align}
where $f_1$ is a scaling function. 
Alternatively, in the steady state $t\to \infty$, by choosing $s^{-\phi}\Delta \omega=1$, we can reduce Eq.~\eqref{eq:powerlaw2_2} to
\begin{align}\label{eq:powerlaw5}
\rho_a(t)=\Delta \omega^{\beta/\phi}f_2( (\Delta \omega)^{-1/\phi} \Delta \kappa)\,,
\end{align}
where $f_2$ is a scaling function. In a steady-state simulation, the absorbing state can be reached because of finite-size effects~\cite{brezuidenhout, sander}. To overcome this problem, when the system reaches the absorbing state, we perform a spontaneous creation, $0 \rightarrow A$.

{In this section, we briefly reviewed the power-law behavior and FSS theory of the absorbing state phase transition. These context will be used in Sec.~\ref{sec:5} to  perform the numerical analysis of the critical exponents.
}
\section{Analytic results}
\label{sec:4}

\subsection{Phase diagram in the mean-field limit}
\begin{figure}
\includegraphics[width=0.85\columnwidth]{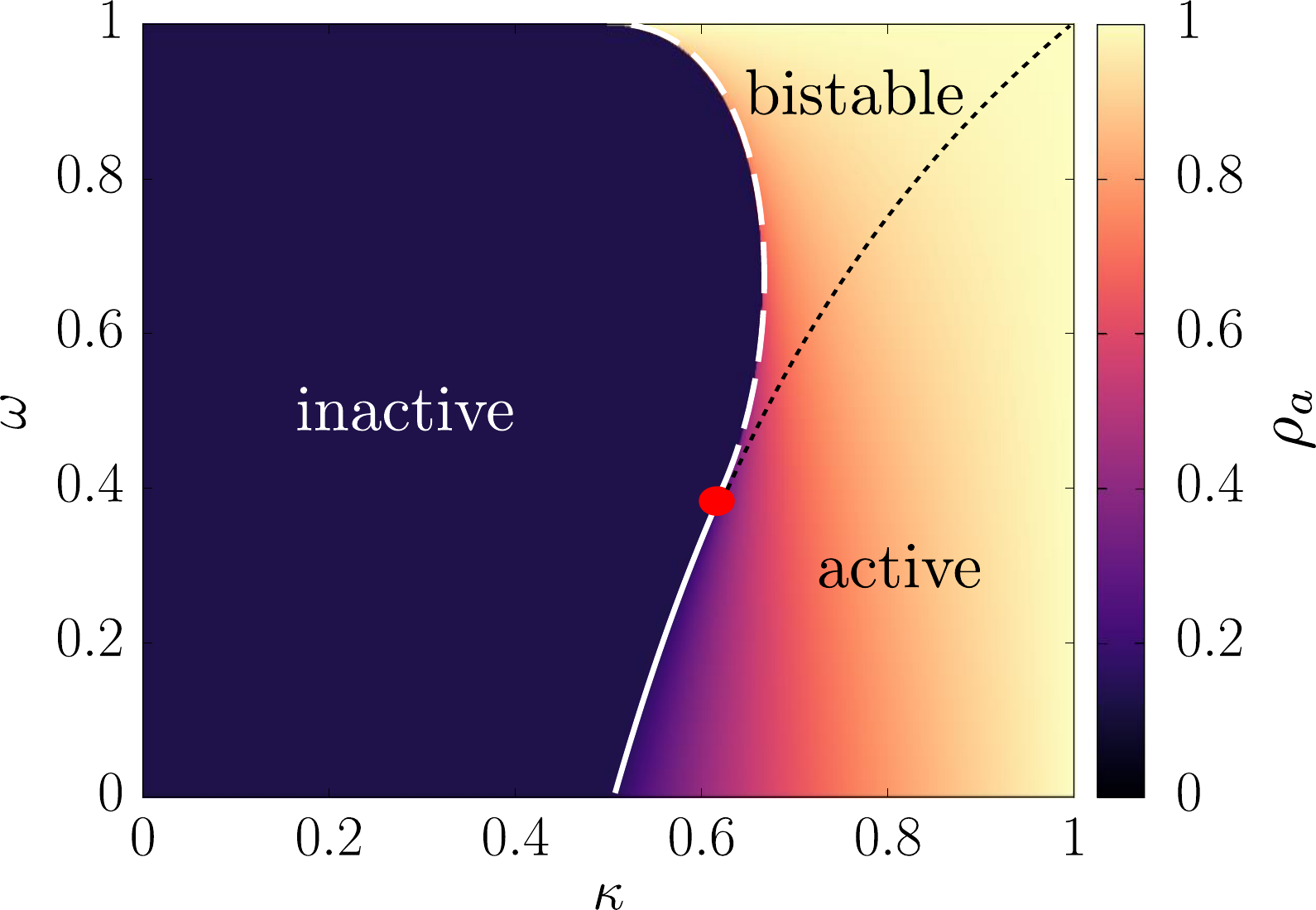}
\caption{Phase diagram of the TCP model in the mean-field limit. A tricritical point (red dot) is located at $(0.6180, 0.3820)$. White solid (dashed) curve represents a continuous (discontinuous) transition.  
\label{fig:fig2}}
\end{figure}
In this section, we recall the analytic result based on the mean-field approach obtained in a previous work~\cite{lastWork}. The density of active sites at time $t$ averaged over the surviving sample is denoted as $\rho_a(t)$. In the mean-field limit, we ignore the effect of local density fluctuations and write the dynamic equation of the LTCP model as 
\begin{align}\label{eq:LTDP_meanfield_homo}
\partial_t \rho_a(t) = -u_2 \rho_a-u_3 \rho_a^2-u_4 \rho_a^3 \,,
\end{align}
where $u_2=\omega\kappa+1-2\kappa$, $u_3=\kappa-\omega-\omega\kappa$, and $u_4=\omega\kappa$. These coefficients are derived on the basis of the reactions listed in Table I.

In the steady state, we set $\partial_t \rho_a=0$ and obtain the solutions as
\begin{align}\label{eq:LTDP_meanfield_homo_2}
\rho_a^*\equiv 0\,~~{\rm and}~~\, \rho^*_{a,\pm}\equiv \frac{-u_3\pm\sqrt{u_3^2-4u_2u_4}}{2u_4}\,.
\end{align}
Linear stability analysis reveals that the first solution, $\rho^*_a=0$, is stable for $u_2>0$ and unstable for $u_2<0$. Thus, $u_2=\omega\kappa+1-2\kappa=0$ is the boundary of the stable solution at the fixed point $\rho^*_a=0$, which is equivalent to the boundary of the active phase in Fig.~\ref{fig:fig2}.

For the second solution, $\rho^*_{a,\pm}$, we analyze the linear stability as 
\begin{align}\label{eq:linearStability}
\delta \dot{\rho}_{a,\pm}&=-(u_2+2u_3\rho^*_{a,\pm}+3u_4\rho^{*2}_{a,\pm})\delta \rho_{a,\pm}\\
&=\rho^*_{a,\pm}(-u_3-2u_4 \rho^*_{a,\pm})\delta \rho_{a,\pm}
=\mp \rho^*_{a,\pm}\sqrt{u_3^2-4u_2u_4}\delta \rho_{a,\pm}\,.
\end{align}
Thus, $\rho^*_{a,+}$ and $\rho^*_{a,-}$ are stable for $\rho^*_{a,+}>0$ and $\rho^*_{a,-} < 0$, respectively. Because {$\rho_a > 0$}, $\rho^*_{a,-}$ is ignored. For $\rho_a=\rho^*_{a,+}$, we obtain two phase boundaries. 
The first is $u_2=0$ and $u_3\ge 0$. Thus, $u_3^2-4u_2u_4\ge 0$. These conditions are rewritten in terms of $(\kappa, \omega)$ as follows: 
\begin{align} \label{eq:critical}
\omega \kappa+1-2\kappa=0 \,,~~{\rm and}~~\omega \le \omega_t\equiv\frac{3-\sqrt{5}}{2} \,.
\end{align}
Thus, $\rho^*_{a,+}=0$. The first equation and second inequality above were used to generate the {white} solid curve in Fig.~\ref{fig:fig2}, and the red dot indicates $\omega=\omega_t$.  

The second phase boundary is obtained from the conditions $u_3^2-4u_2u_4=0$ and $u_3\le0$. These conditions lead to 
\begin{equation}\label{eq:firstorder}
(\kappa-\omega-\omega \kappa)^2-4\omega \kappa(1-2\kappa+\omega \kappa)=0
\,\, {\rm for} \,\,
\omega \ge \omega_t
\,\, {\rm and }\,\,
 \kappa\ge 0.5,
\end{equation}
where $\rho^*_{a,+} \ge 0$. This phase boundary is drawn as a white dashed curve in Fig.~\ref{fig:fig2}.

There exist three phases in the phase diagram (Fig.~\ref{fig:fig2}): i) the inactive (absorbing) phase with $\rho_a=0$, ii) the active phase with $\rho_a=\rho^*_{a,+} >0$, and iii) the bistable phase with two stable fixed points, $\rho_a=0$ and $\rho_a=\rho^*_{a,+} > 0$. The phase boundaries are determined by the conditions derived above. We will show later that the phase transition across the first boundary above (indicated by the {white} solid curve) is second-order, whereas that across the second boundary (indicated by the dashed curve) is first-order. Therefore, a tricritical point is formed at $(\kappa_t, \omega_t)$. The critical exponent of the order parameter defined as $\rho_a\sim (\kappa-\kappa_c)^{\beta}$ across the white solid curve is found to be $\beta=1$ for $\omega<\omega_t$, and the exponent $\beta_t$ for $\rho_a\sim (\kappa-\kappa_t)^{\beta_t}$ is found to be $\beta_t=1/2$ at $\omega=\omega_t$. 

To confirm our analytic result, we numerically verified the phase diagram on the fully connected lattice. {Specifically, using the FSS theory, we obtained the tricritical point and critical exponents presented in the next subsection corresponding to the analytic results.}

\subsection{Hyperscaling relation for LTDP}
\label{sec:hyperScaling}

In this section, we recall the field-theoretic analysis performed in the previous work~\cite{lastWork} to obtain the exact scaling relation and mean-field exponents.
To account for the spatial fluctuations and noise induced by active particles occupying active sites, we set up the Langevin equation as follows: 
\begin{align}\label{eq:LTDP_meanfield_inhomo}
\partial_t \rho_a &= D_{\sigma}\nabla^{\sigma} \rho_a+D\nabla^{2}\rho_a-u_2 \rho_a-u_3 \rho_a^2-u_4 \rho_a^3+\xi\,, 
\end{align}
where $D_{\sigma}$ and $D$ are the diffusion constants obtained from a small momentum expansion, which are given by $(1-\omega)\kappa\int d\bm{r}' P(|\bm{r}-\bm{r}'|)\rho_a(\bm{r}')\approx (1-\omega)\kappa \rho_a+D_{\sigma}\nabla^{\sigma} \rho_a+D\nabla^{2}\rho_a$. 
The noise $\xi(\bm{r}, t)$ is a multiplicative Gaussian random variable with zero mean and a correlation of 
\begin{align}
\langle \xi(\bm{r}, t)\xi(\bm{r}, t) \rangle = \Gamma \rho_a(\bm{r},t) \delta^d(\bm{r}-\bm{r}')\delta(t-t')\,.
\end{align}
Using the Martin--Siggia--Rose--Janssen--de Dominicis formalism~\cite{MSRJD1,MSRJD2,MSRJD3,MSRJD4,MSRJD5} for the Langevin equation, we obtain the action as follows:
\begin{align} \label{eq:action}
S =\int d{\bm x} \, {\rho_a^\prime}\left[
\partial_t  - D \nabla^2 - D_{\sigma}\nabla^{\sigma} +u_2 
+ u_3\rho_a + u_4 \rho_a^2 -\frac{\Gamma}{2} \rho_a \right] \rho_a \,,
\end{align}
where $\rho_a^\prime$ is an auxiliary field, and ${\bm x} = ({\bm r}, t)$. 

If $u_3$ is finite, $u_4$ is irrelevant at $d_c$, which implies that the action described by Eq.~\eqref{eq:action} belongs to the long-range DP (LDP) class.
It is satisfied by the so-called rapidity-reversal (or duality) symmetry, which is invariant under the exchange $\rho_a(\bm{r},t)\leftrightarrow -\rho_a^\prime(\bm{r},-t)$.
Rapidity-reversal symmetry implies that the critical exponents $\beta$ and $\beta^\prime$ must be identical.
It was revealed that in the LDP class, $D_\sigma$ is not renormalized~\cite{LDP1,LDP2}.
This means that $D_\sigma$ is invariant under the scaling transformation; thus, one obtains the exact scaling relation $d+z-\sigma-2z\delta=0$~\cite{LDP1,LDP2,LDP3,LDP4,LDP5}.

At the tricritical point, $u_3=0$, the rapidity-reversal symmetry is broken, and $\beta \neq \beta^\prime$. Crossover behavior occurs when more than one fixed point appears in the phase diagram. 
Scaling theory is used to obtain the mean-field critical exponents: 
\begin{align}
\beta=0.5\,,\; \beta^\prime=1\,,\; \nu_{\bot}=1/\sigma\,,\; \nu_{\|}=1\,,\; z=p\,,\;\phi=0.5\,,
\label{eq:exponentsTDP}
\end{align}
which are expected to be valid above the upper critical dimension $d_c=1.5\sigma$.
For the LDP class, loop corrections can be represented as an integer power series in momentum space~\cite{LDP2}. This can be applied to LTDP as well, which means that the coefficient of the fractional Laplacian is not renormalized (see Appendix~\ref{appendixB}). This implies that the coefficient of the fractional Laplacian must be invariant under the renormalization group (RG) transformation.
Hence, one can obtain the so-called hyperscaling relation
\begin{align}
d+z-\sigma-z(\delta+\delta^\prime)=0 \,,
\label{eq:hyperscaling}
\end{align}
which is valid below the upper critical dimension, $d\le d_c$. 

Below $d_c$, the universal features of the LTCP model depend on $\sigma$, which is within the interval [$\sigma_{c1}$,$\sigma_{c2}$]. Thus, we consider the following three domains. First, below $\sigma_{c1}$, the interaction range can be superdiffusive; thus, mean-field critical behavior appears. {In other words, above the upper critical dimension $d>d_c=1.5\sigma$, mean-field  behavior is expected.} Thus, $\sigma_{c1}=2d/3$. Second, {$\sigma_{c2}$ is} determined {to be 1.36067} using Eq.~\eqref{eq:hyperscaling}. Finally, in the regime $\sigma > \sigma_{c2}$, the exponents {are reduced to} those of the STDP class.

\section{Numerical Results}
\label{sec:5}

\subsection{STCP model in two dimensions}
\label{sec:5-2}
\begin{figure}[ht!]
\includegraphics[width=0.85\columnwidth]{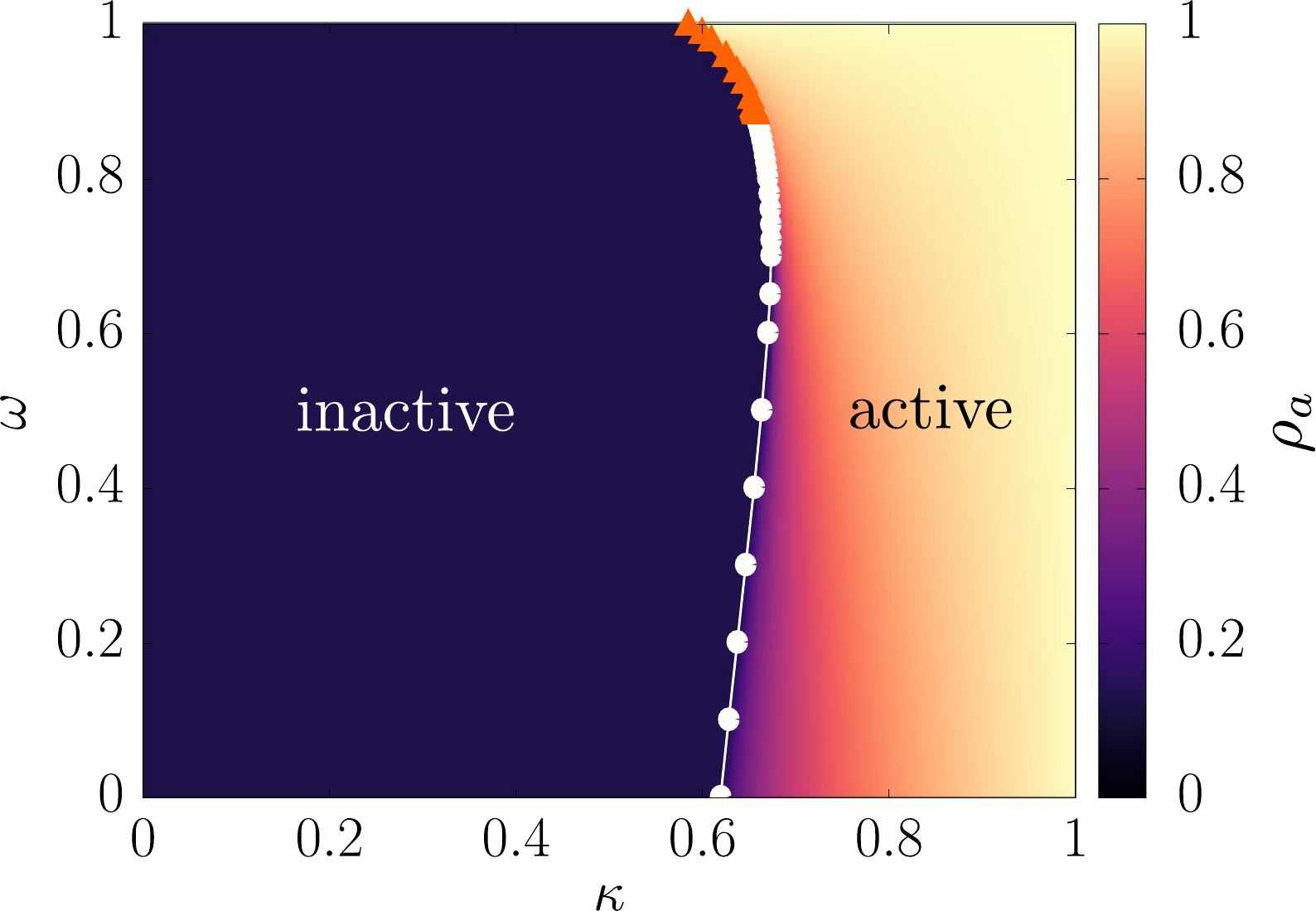}
\caption{Phase diagram of the $m$-TCP model in two dimensions. A tricritical point is located at $(0.6606466,0.879)$. White (Orange) curve represents a continuous (discontinuous) transition. At $\omega=0$, the model is reduced to the CP model at $\kappa_c=0.622466$. The data points (white circles and orange triangles) represent numerical results.
\label{fig:fig3}}
\end{figure}

\begin{figure}[ht!]
\includegraphics[width=0.85\columnwidth]{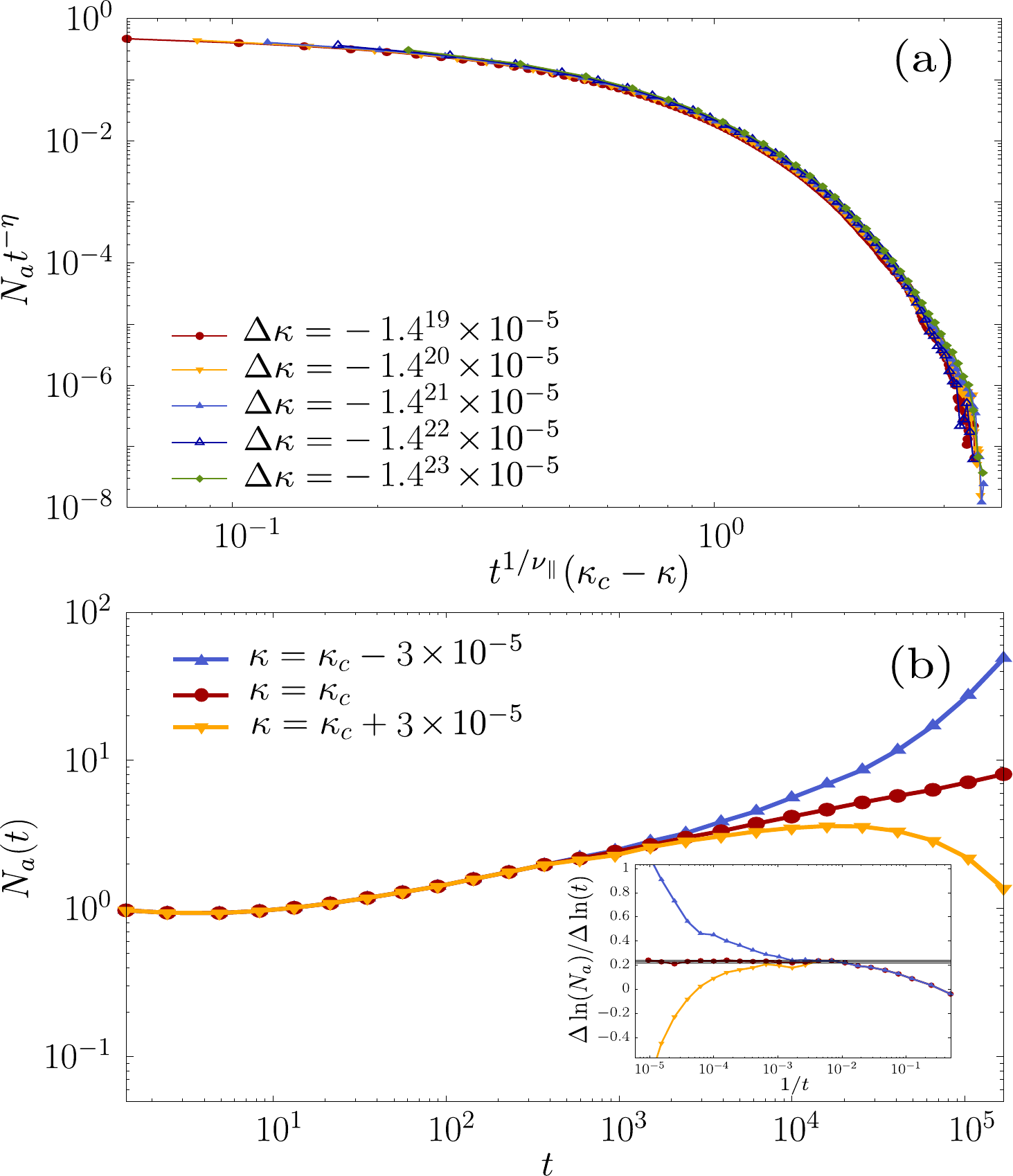}
\caption{(a) Scaling plot of $N_at^{-\eta}$ versus $t^{1/\nu_{\|}}(\kappa_c-\kappa)$ for different values of $\kappa$. Data points collapse well onto a single
curve for $\eta=0.230$ and $\nu_{\|}=1.295$. 
{(b) Plot of {$N_a(t)$} versus $t$ at and around $\kappa_c$. Inset: Local slopes of $N_a(t)$ versus $1/t$ for these data points obtained in (a). $\omega=0.6$, and $\kappa_c=0.67326$.}
\label{fig:fig4}}
\end{figure}

\begin{figure}[ht!]
\includegraphics[width=.85\columnwidth]{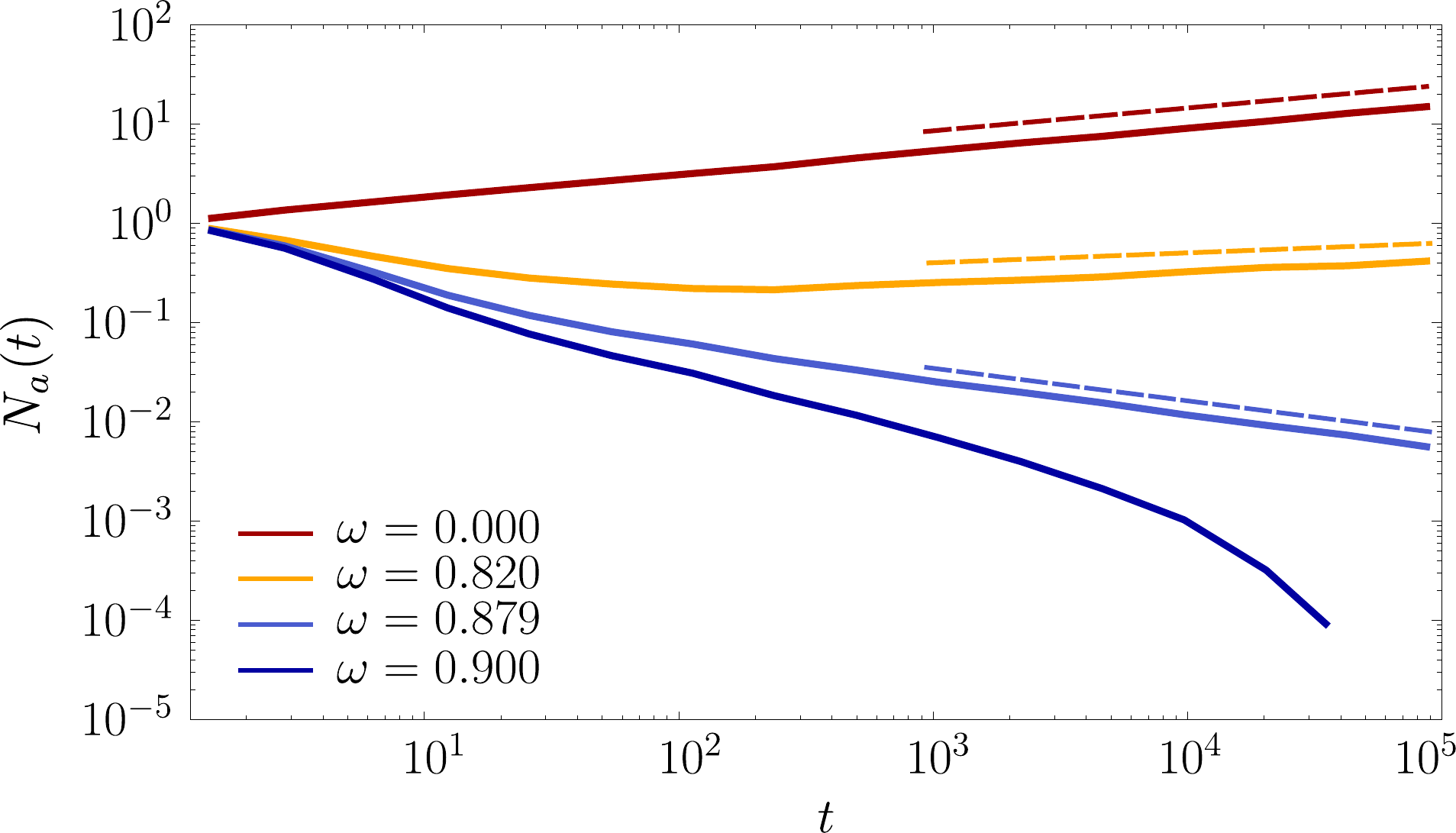}
\caption{Plot of $N_a(t)$ for different values of $\omega$: $\omega=0$ at the DP point; $\omega=0.879$ at the TDP point; $\omega=0.82$ in the crossover region between these two points; and $\omega=0.9$ in the first-order transition domain. Dashed lines are guidelines with slope $0.230, 0.102,$ and $-0.353$, from the top. The system size is taken as $N=10^8$.} 
\label{fig:fig5}
\end{figure}

\begin{figure*}[ht!]
\includegraphics[width=1.75\columnwidth]{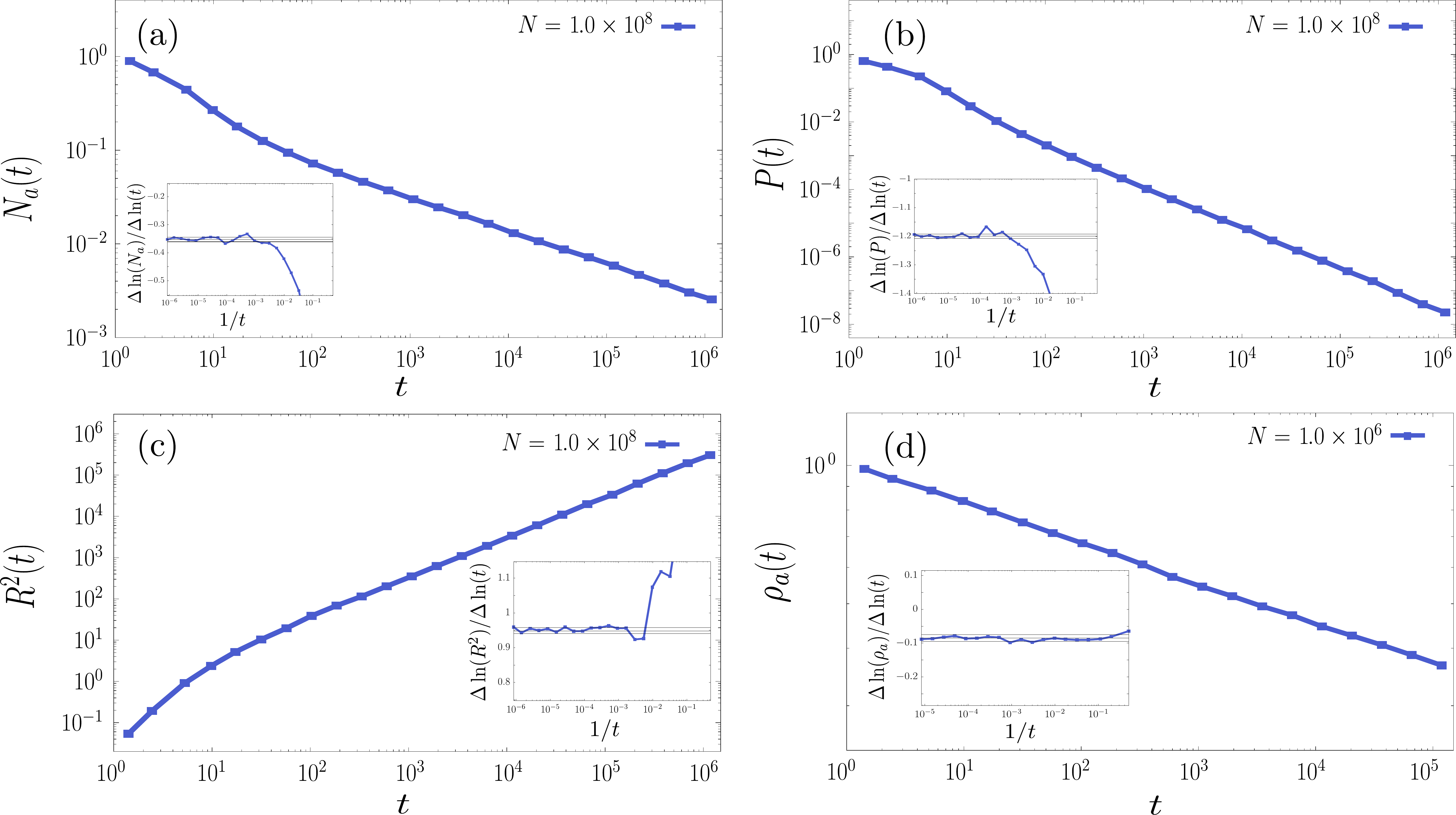}
\caption{{Plots of four physical quantities used to characterize the absorbing transition of the $m$-TCP model in two dimensions at a tricritical point: (a) $N_a(t)$ versus $t$, (b) $P(t)$ versus $t$, (c) $R^2(t)$ versus $t$, and (d) $\rho_a(t)$ versus $t$. The exponent values are estimated as follows: (a) $\eta= -0.35\pm 0.008$, (b) $\delta^\prime = 1.22\pm 0.008$, (c) {$2/z=0.947\pm 0.004$}, and (d) $\delta=0.09\pm 0.01$. Insets: Local slopes of each quantity versus $1/t$ to confirm the estimated slopes.} 
\label{fig:fig6}}
\end{figure*}

\begin{figure}[ht!]
\includegraphics[width=0.85\columnwidth]{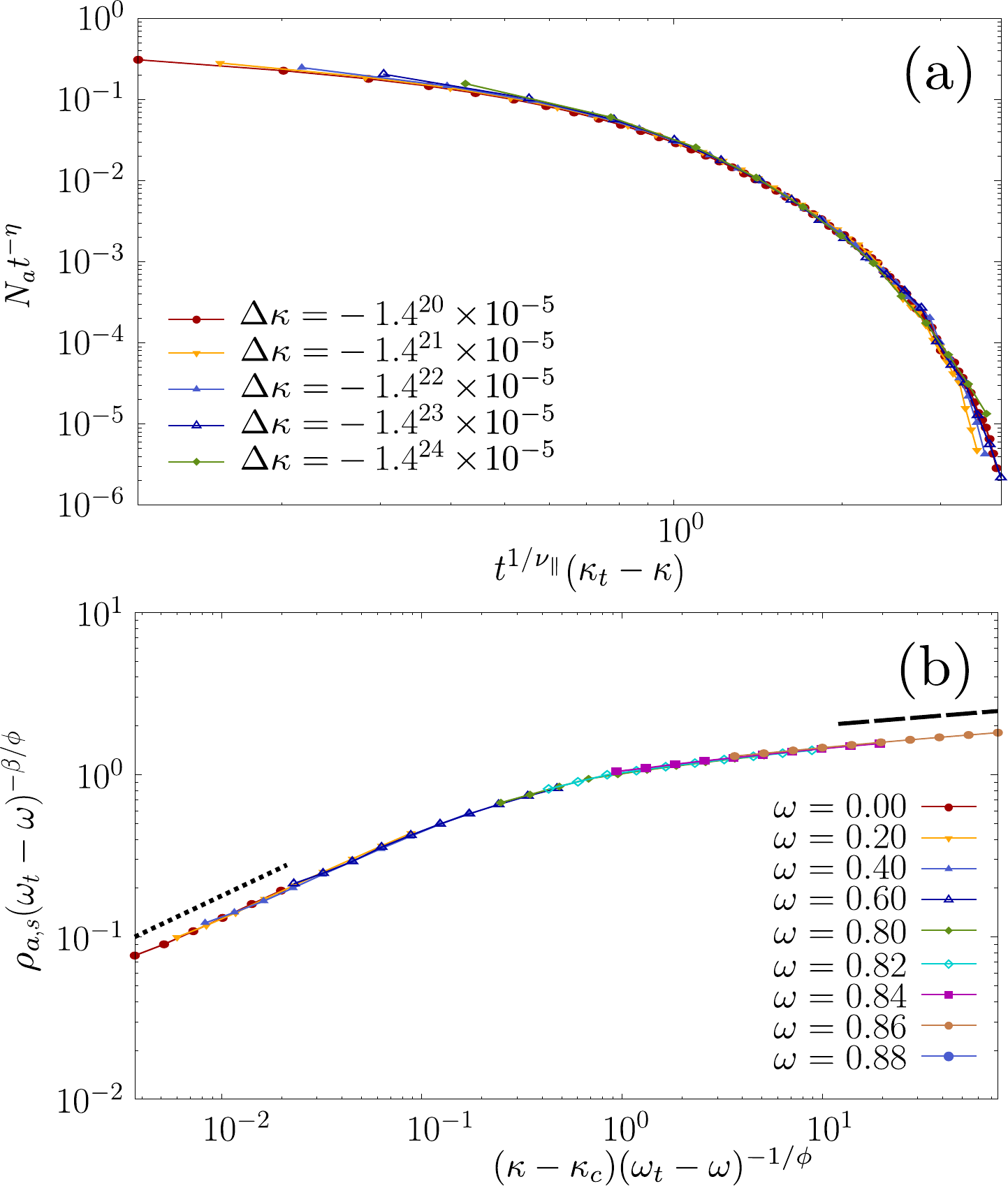}
\caption{FSS analysis of the $m$-TCP model. (a) Scaling plot of $N_at^{-\eta}$ versus $t^{1/\nu_{\|}}(\kappa_t-\kappa)$ for different values of $\kappa$. Data points collapse well onto a single curve for $\kappa_t=0.6606466$, $\eta=-0.353$, and $\nu_{\|}=1.16$. 
(b) Scaling plot of $\rho_{a,s}(\omega_t-\omega)^{-\beta/\phi}$ versus $(\kappa-\kappa_c)(\omega_t-\omega)^{-1/\phi}$ for different values of $\omega$, where $\rho_{a,s}$ represents $\rho_a$ in the steady state. Dotted (Dashed) line is a guideline with slope $\beta_{\rm DP}=0.584$ ($\beta_{t}=0.101$). Data points collapse well onto a single curve for $\phi=0.52$. 
} 
\label{fig:fig7}
\end{figure}

\begin{figure}[ht!]
\includegraphics[width=0.85\columnwidth]{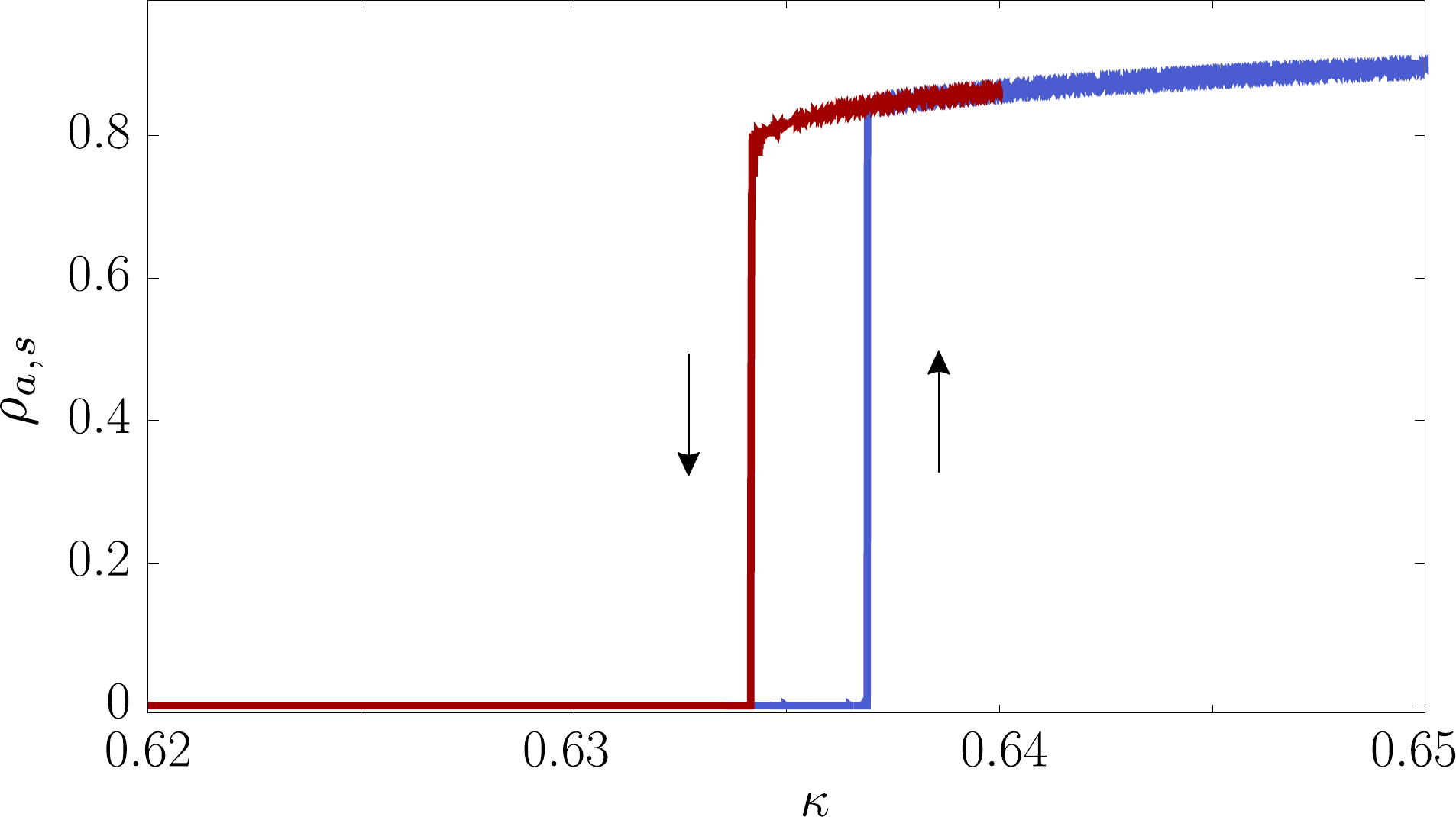}
\caption{Plot of $\rho_{a,s}$ versus $\kappa$ for the $m$-TCP model at a fixed $\omega = 0.95>\omega_t$. A hysteresis curve is obtained. The system size is $N=10^6$. 
}
\label{fig:fig8}
\end{figure}

\begin{table*}[ht!]
\caption{2$d$ TDP universality at tricritical point $(\kappa_t,\omega_t)$ for various models.
Here, we determine the tricritical point of L\"ubeck's TCP model by finding the power-law behavior of the number of active sites $N_a(t)$.
}
\begin{center}
\setlength{\tabcolsep}{7pt}
{\renewcommand{\arraystretch}{1.5}
\begin{tabular}{ccccccccccc}
    \hline
    \hline
    Model & $(\kappa_t,\omega_t)$ & $\nu_{\|}$ & $z$ & $\delta$ & $\delta^\prime$ & $\eta$ \\
    \hline
    Generalized Domany$-$Kinzel ~\cite{grassberger} & $(0.1813672, 2.795)$ & $1.156(4) $ & $2.110(6) $ & $0.087(3)$  & $1.218(7) $  & $-0.353(9)$\\
    Ordinary TCP~\cite{lubeck} & $(0.286237, 0.919)$ & $1.15\pm 0.005 $ &$2.11\pm 0.01$ & $0.09\pm 0.01$  & $1.22\pm 0.008$  & $-0.35\pm 0.008$\\
    Modified TCP & $(0.6606466, 0.879)$ & $1.15\pm 0.005 $ & $2.11\pm 0.01 $ & $0.09\pm 0.01$  & $1.22\pm 0.008$  & $-0.35\pm 0.008$\\
    \hline
    \hline
\end{tabular}}
\label{tab:tab2}
\end{center}
\end{table*}

We consider an STCP model {called} the $m$-TCP model to distinguish it from other previous models designed as models of the STCP class. This model is a simple version of {the} LTCP model obtained by replacing the long-range interaction with short-range interaction, which we will consider next. In fact, the STCP model was explored in Refs.~\cite{lubeck,grassberger,windus} using slightly different rules, but the numerical values of their critical exponents differed from each other. The origin of this difference will be discussed later. Here, we check the justification for our LTCP model using the simplified version, the $m$-TCP model, by comparing our simulation results with those obtained in Refs.~\cite{lubeck,grassberger,windus}. This $m$-TCP model contains two control parameters, $\kappa$ and $\omega$. In $(\kappa,\omega)$ space, there exist second-order and first-order phase transition curves and a tricritical point at which the two transition curves meet.

To determine the second-order curve, we find a critical point $\kappa_c$ for each value of $\omega$ in the region $\omega < \omega_t$ as follows. First, we use the FSS method based on Eq.~\eqref{eq:powerlaw3}. We take the scale factor $s$ as $s=\kappa_c-\kappa$ and plot $N_a(t)t^{-\eta}$ versus $(\kappa_c-\kappa)t^{1/\nu_{\|}}$. If we choose $\kappa_c$ correctly, then the data points for different $\kappa$ values would collapse onto a single curve. Indeed, we obtain this result, for instance, for $\omega=0.6$ with $\kappa_c=0.67326$ [Fig.~\ref{fig:fig4}(a)]. In the second method, we check the local slope of the curve of {$N_a(t)$} as a function of $t$. If we choose $\kappa_c$ correctly, then {$N_a(t)$} {would} exhibit power-law behavior as a function of $t$ with the exponent {$\eta(\omega)$} [Fig.~\ref{fig:fig4}(b)].
{Using} these two methods, we determine the critical points $\kappa_c$ for each value of $\omega$. 
 
We obtain the phase diagram shown in Fig.~\ref{fig:fig3}. When $\omega=0$, the absorbing transition belongs to the DP class, and thus $\eta\approx 0.230$. We trace the value of the exponent $\eta$ as a function of $\omega$ in Fig.~\ref{fig:fig5}. 
The $\omega$ {values are} chosen as follows:
(i) $\omega=0$ (DP class);
(ii) $\omega=0.82$ (in the crossover region from DP to TDP);
(iii) $\omega=0.879$ (TDP class); and 
(iv) $\omega=0.9$ (in the region of the first-order transition).
In Fig.~\ref{fig:fig5}, there are two generic power-law lines at $\omega=0$ and $\omega=0.879$.
At the tricritical point, we obtain the tricritical exponents as $\eta = -0.35 \pm 0.008$, $\delta^\prime=1.22\pm 0.008$, $z=2.11\pm 0.01$, and $\delta=0.09 \pm 0.01$ in Fig.~\ref{fig:fig6}.
{When we perform the data collapse, the error bars are measured by controlling the exponents until the data collapse breaks down.}
The exponent $\nu_{\|}$ is obtained from the rescaling plot of $N_a(t)t^{-\eta}$ versus $t^{1/\nu_{\|}}(\kappa_t-\kappa)$ for different $\kappa$ values in Fig.~\ref{fig:fig7}(a). 
In Fig.~\ref{fig:fig7}(b), the crossover exponent $\phi$ is obtained from the rescaling plot of $\rho_{a,s}(\omega_t-\omega)^{-\beta/\phi}$ versus $(\kappa-\kappa_c)(\omega_t-\omega)^{-1/\phi}$. In this case, $\phi=0.52\pm 0.02$ is obtained, in agreement with the result in Ref.~\cite{lubeck}.  
We remark that the authors of Refs.~\cite{lubeck,grassberger,windus} considered TCP models with slightly different reaction rules. Further, they obtained slightly different critical exponent values. It was argued that this discrepancy results from the different methodologies used to determine the tricritical point in Ref.~\cite{lubeck}. The author of Ref.~\cite{lubeck} used FSS of the order parameter in the steady state. It is difficult to find a tricritical point correctly using this method, because FSS in the steady state is not sensitive to $\kappa_t$. On the basis of our two criteria, we obtain $\omega_t=0.9190$ instead of the value of $0.9055$ in Ref.~\cite{lubeck}. At our tricritical point, we obtain critical exponent values similar to those in Ref.~\cite{grassberger}. In Table~\ref{tab:tab2}, we list the three sets of critical exponent values of the STCP model obtained using three different rules.

For $\omega>\omega_t$, as represented by the orange curve in Fig.~\ref{fig:fig3}, a first-order transition occurs. One of the features of the first-order transition is the presence of a hysteresis curve.
Thus, we check whether a hysteresis curve is indeed generated. After taking an $\omega$ value larger than $\omega_t$, say $\omega=0.95$, we calculate the LTDP dynamics for a given $\kappa$ and obtain $\rho_a(\kappa)$ in the steady state. Next, we increase $\kappa$ slightly and simulate the LTDP dynamics again; we obtain $\rho_a$ in the steady state. We repeat this process in the forward direction, in which $\kappa$ is increased, and in the backward direction, in which $\kappa$ is decreased. Indeed, we obtain a hysteresis curve, as shown for $\omega=0.95$ in Fig.~\ref{fig:fig8}. 
Here, we determine the critical point of the first-order transition {following the method used in Refs.~\cite{brosilow,lubeck}.} For fixed $\omega$ and $\kappa$, we set up an initial configuration in which half of the sites are assigned to the active state and the remaining sites are assigned to the inactive state, and the LTDP dynamics is simulated. The system reaches either the absorbing state ($\rho_a=0$) or the active state ($\rho_a>0$) depending on initial configuration and given $\kappa$. We measure the fraction of initial configurations that reach the absorbing state as a function of $\kappa$. The transition point $\kappa_c$ is determined as the one at which the fraction becomes half. 

\subsection{LTCP model in two dimensions}
\label{sec:5-3}

\begin{figure}[ht!]
\includegraphics[width=0.85\columnwidth]{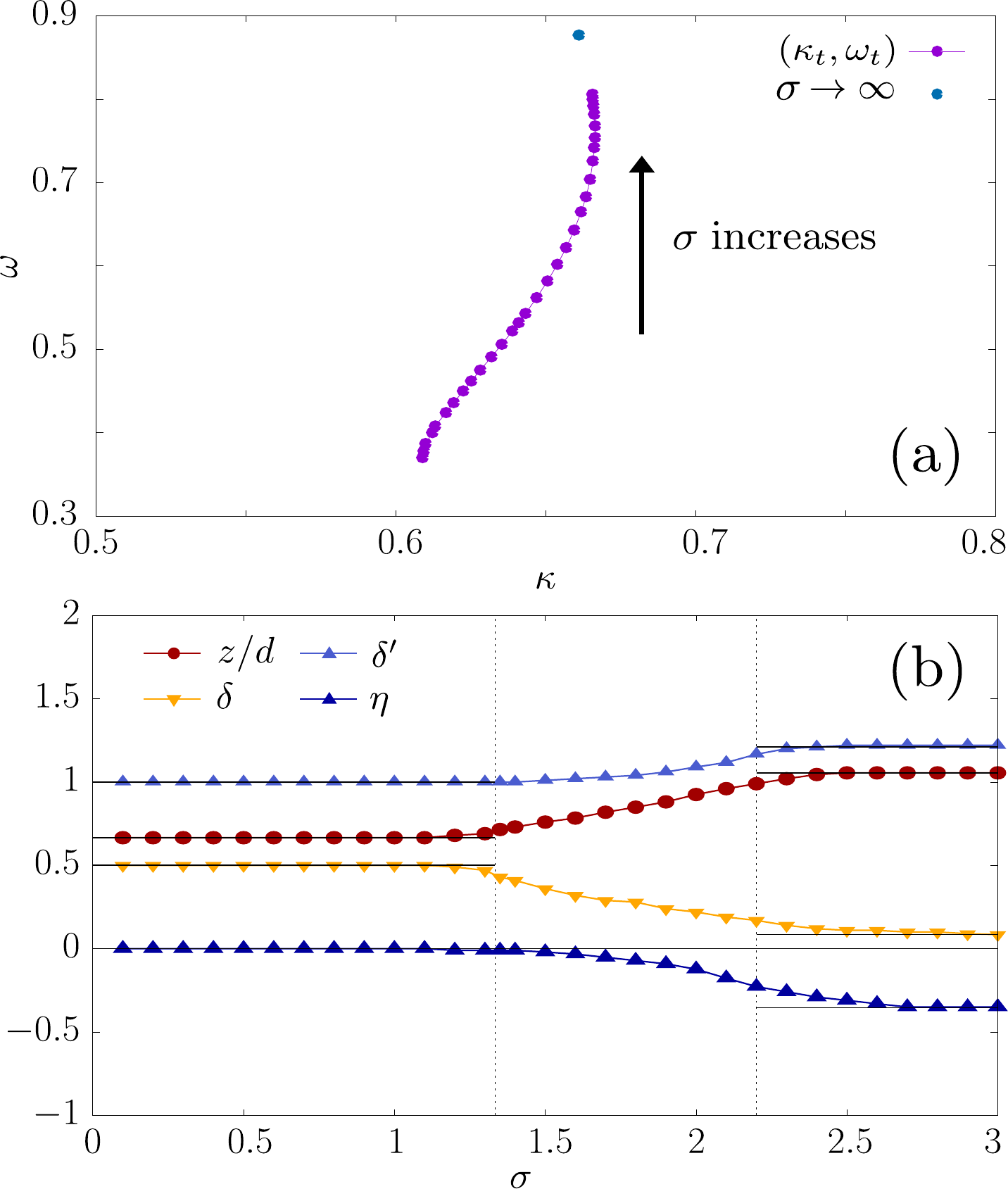}
\caption{For the LTCP model in two dimensions, (a) plot of the tricritical points in ($\kappa,\omega$) space for different $\sigma$ values in $[0.1,3.0]$. (b) Plots of the critical exponents $z/d$, $\delta$, $\delta^\prime$, and $\eta$ as a function of $\sigma$. $\sigma_{c1}$ and $\sigma_{c2}$ are indicated by vertical dotted lines. The thin solid lines in the regions $\sigma<\sigma_{c1}$ and $\sigma > \sigma_{c2}$ are guidelines showing that the curves converge to constant values.} 
\label{fig:fig9}
\end{figure}

\begin{figure*}[ht!]
\includegraphics[width=1.75\columnwidth]{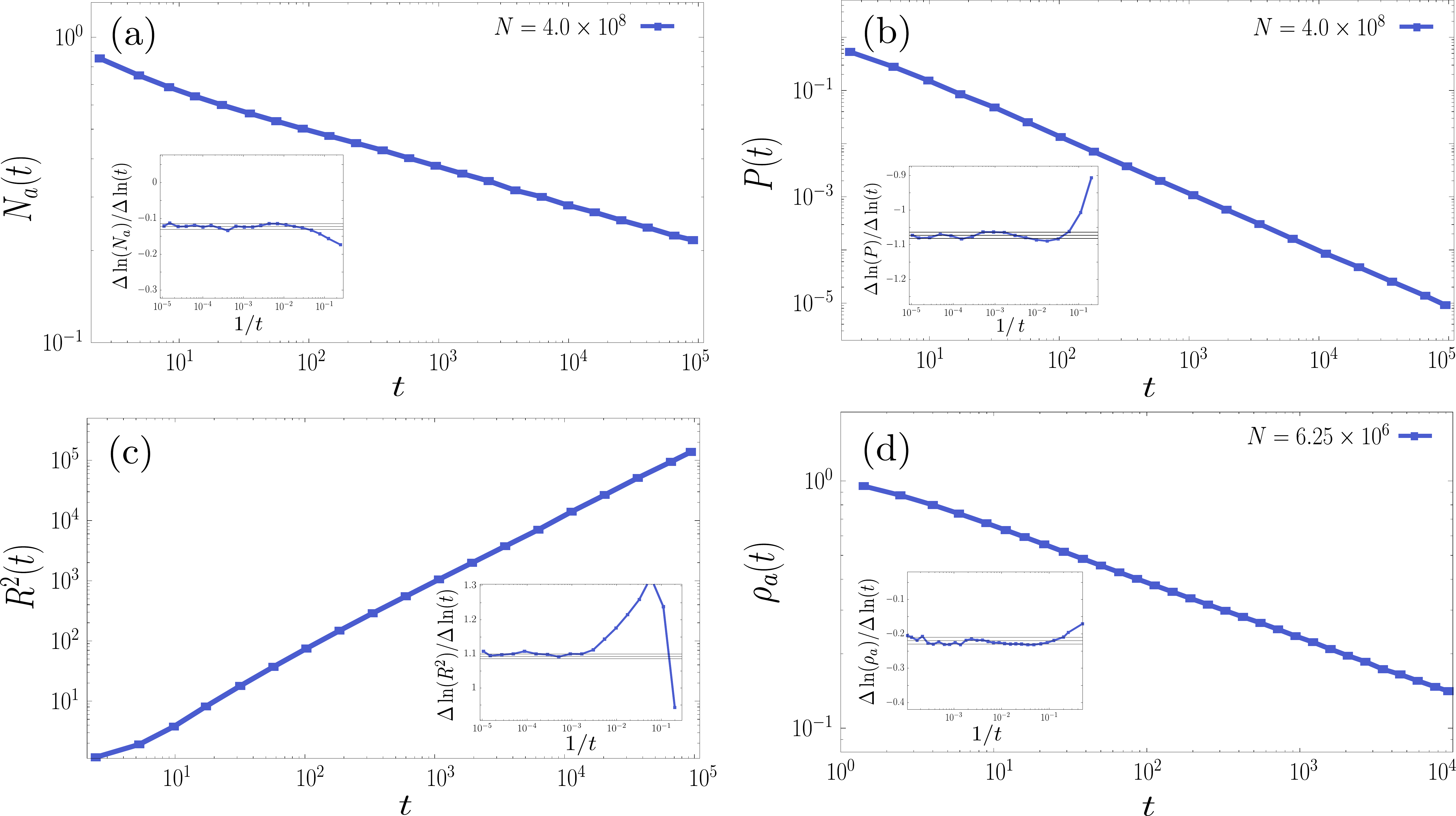}
\caption{For the LTDP model with $\sigma=2.0$ in two dimensions, plots of (a) $N_a(t)$, (b) $P(t)$, (c) $R^2(t)$, and (d) $\rho_a(t)$ versus $t$. We obtain the exponent values as $\eta=-0.129\pm 0.010$, $\delta^\prime=1.073\pm 0.010$, $2/z=1.087\pm 0.010$, and $\delta=0.212\pm 0.010$, respectively. Insets: local slopes of each quantity versus $1/t$. 
\label{fig:fig10}}
\end{figure*}

\begin{figure}[ht!]
\includegraphics[width=0.85\columnwidth]{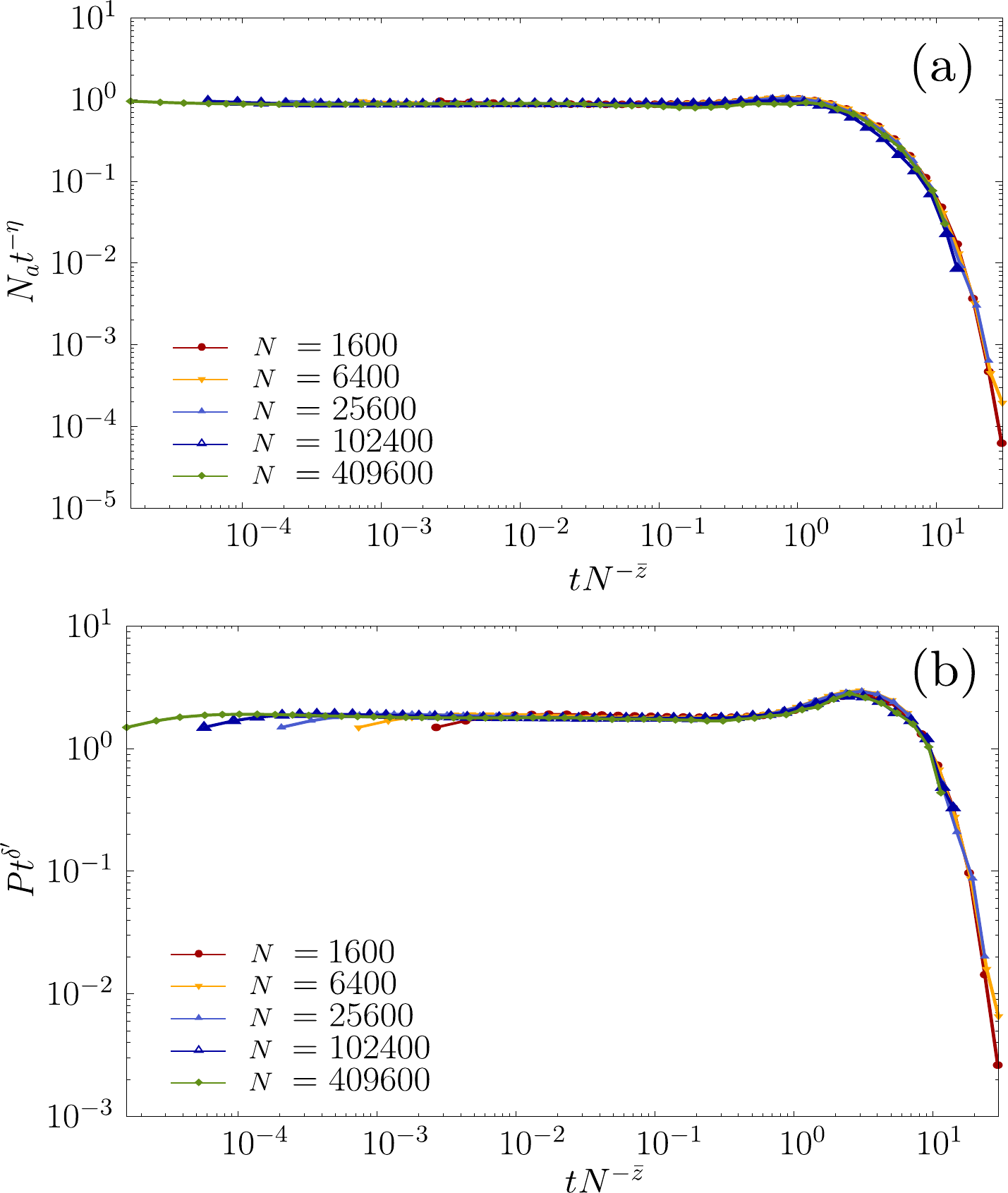}
\caption{For the LTDP model with $\sigma=2.0$ in two dimensions, (a) scaling plot of $N_at^{-\eta}$ versus $tN^{-\bar{z}}$ for $\eta=-0.129$ and $\bar{z}=0.922$. (b) Scaling plot of $P(t)t^{\delta^\prime}$ versus $tN^{-\bar{z}}$ for $\delta^\prime=1.073$ and $\bar{z}=0.922$. 
\label{fig:fig11}}
\end{figure}

\begin{figure}[ht!]
\includegraphics[width=0.85\columnwidth]{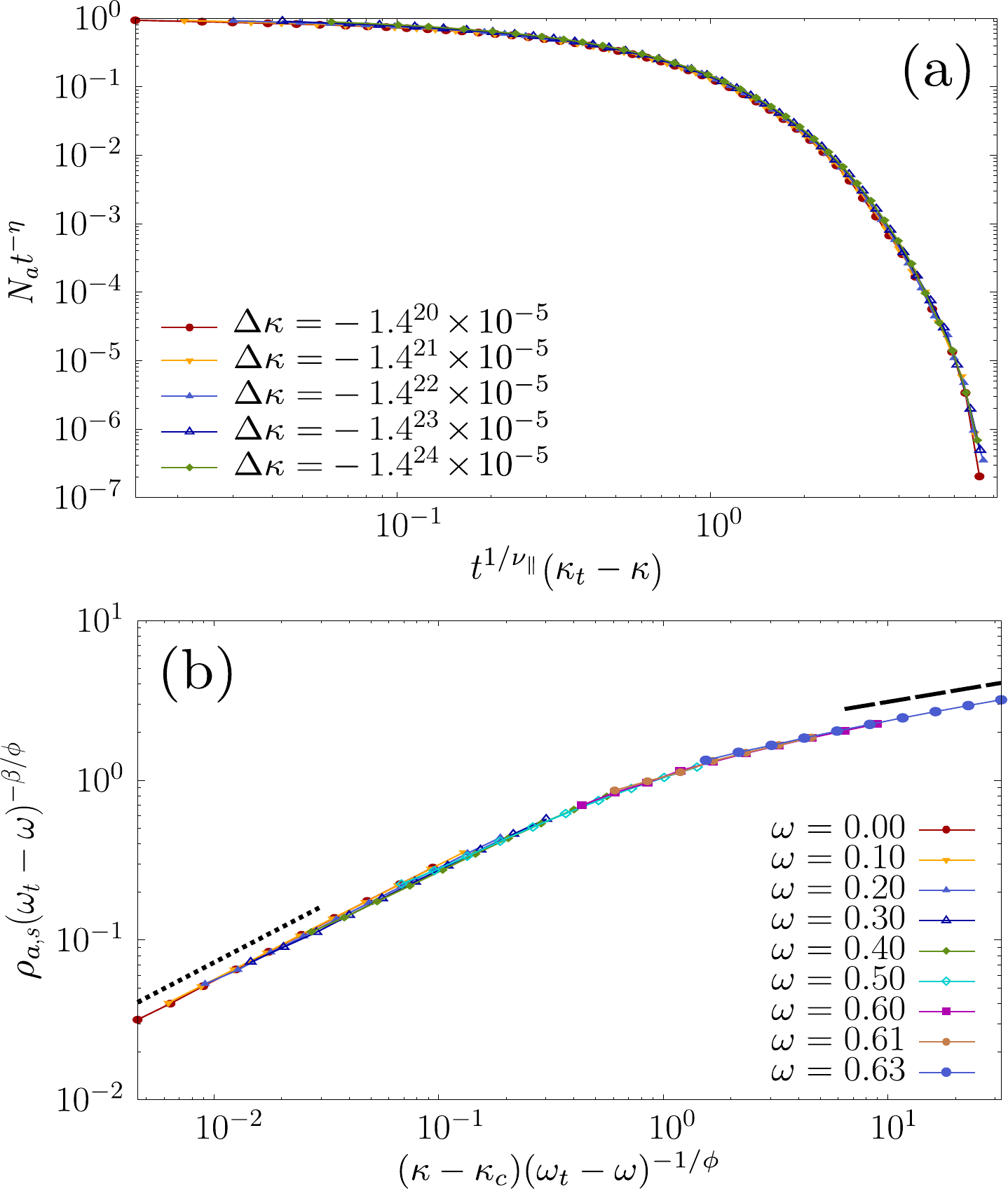}
\caption{For the LTDP model with $\sigma=2.0$ in two dimensions, (a) scaling plot of $N_at^{-\eta}$ versus $t^{1/\nu_{\|}}(\kappa_t-\kappa)$ for different values of $\kappa$. Data points collapse well onto a single curve for {$\kappa_t=0.661663$}, $\eta=-0.129$, and $\nu_{\|}=1.07$. (b) Scaling plot of $\rho_{a,s}(\omega_t-\omega)^{-\beta/\phi}$ versus $(\kappa-\kappa_c)(\omega_t-\omega)^{-1/\phi}$ for different values of $\omega$.  Dotted (Dashed) line is a guideline with slope $\beta_{\rm LDP}=0.7316$ ($\beta_{t}=0.2236$). Data points collapse well onto a single curve for $\phi=0.52$.}
\label{fig:fig12}
\end{figure}

\begin{figure}[ht!]
\includegraphics[width=0.85\columnwidth]{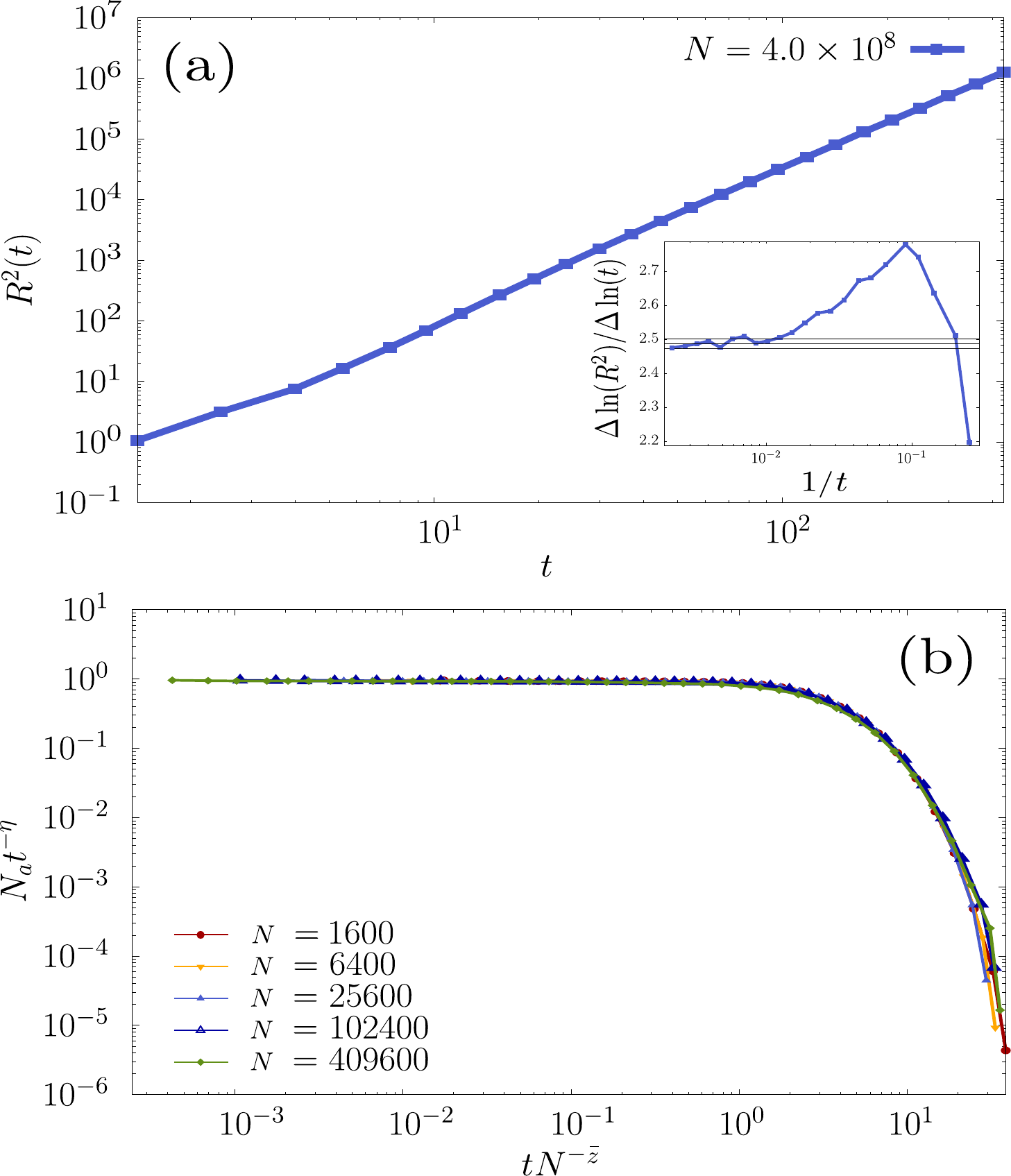}
\caption{{Plots of LTCP in two dimensions for $\sigma=0.8$. (a) Plot of $R^2(t)$ versus $t$. Inset represents local slopes of each quantity versus $1/t$. (b) Scaling plot of $N_at^{-\eta}$ versus $tN^{-\bar{z}}$ for $\eta=0$ and $\bar{z}=0.666$. 
We obtain the exponent values as (a) $2/z=2.491\pm 0.010$ and (b) $\bar{z}=0.666\pm 0.003$.}
}
\label{fig:fig13}
\end{figure}

\begin{figure}[ht!]
\includegraphics[width=0.85\columnwidth]{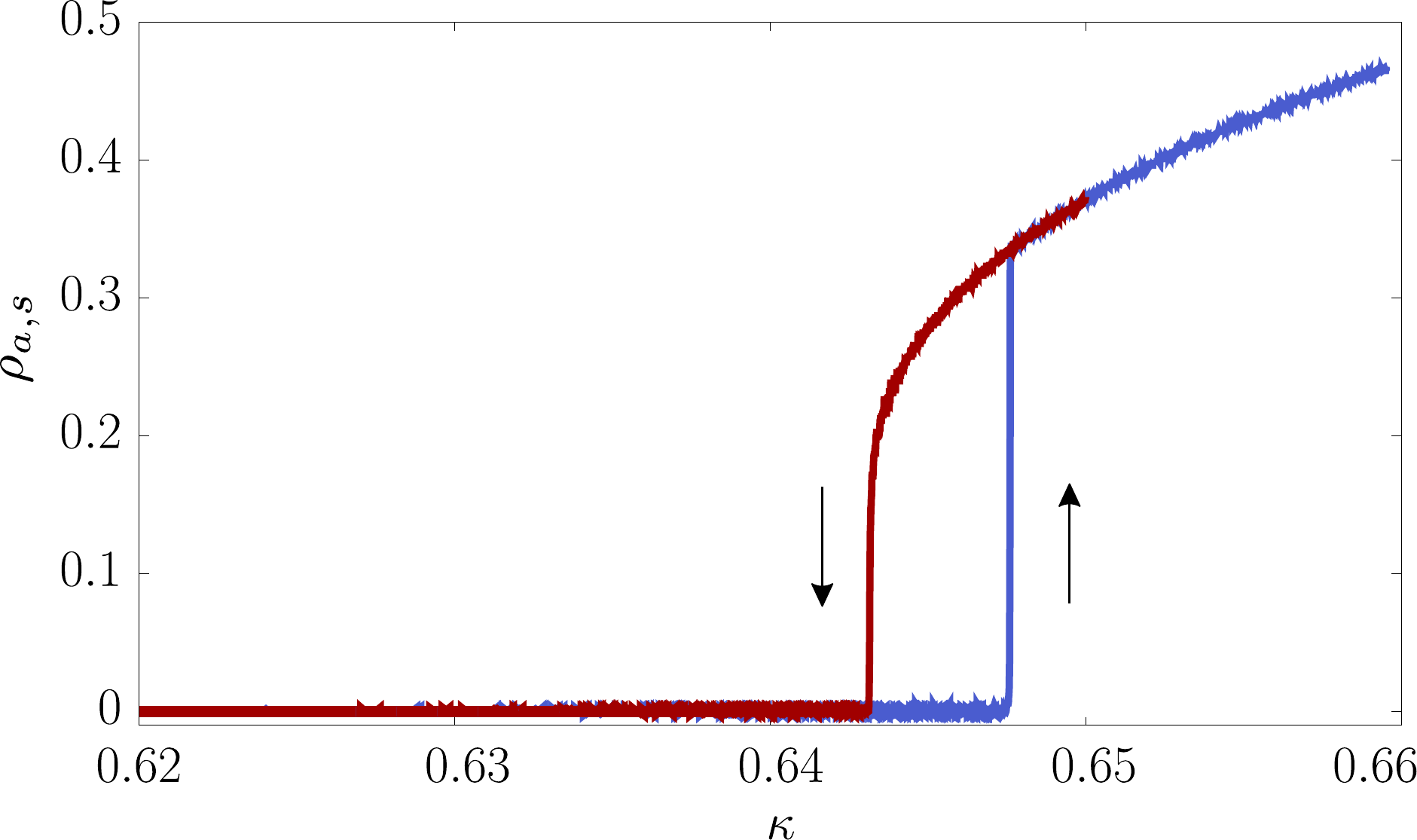}
\caption{For the LTCP model with $\sigma=1.0$ in two dimensions at $\omega=0.55 > \omega_t$, plot of $\rho_{a,s}$ versus $\kappa$. A hysteresis curve is obtained. The system size is $N=10^6$.
}
\label{fig:fig14}
\end{figure}

We perform numerical simulations of the LTCP model in two dimensions, in which the long-range interaction exponent $\sigma$ is varied in the range $[0.1,\,3.0]$ in steps of $\Delta \sigma=0.1$. For each value of $\sigma$, we determine both the critical points $(\kappa_c,\,\omega_c)$ and the tricritical point $(\kappa_t,\,\omega_t)$ using the two methods employed in the previous subsection. As in the phase diagram of the $m$-TCP model, a second-order (first-order) transition occurs for $\omega < \omega_t(\sigma)$ ($\omega > \omega_t(\sigma)$). Thus, a tricritical point appears for each value of $\sigma$, as shown in Fig.~\ref{fig:fig9}(a). 
The second-order transition belongs to the long-range DP class when $\omega \ll \omega_t(\sigma)$ for the given $\sigma$ values. However, as $\omega$ approaches $\omega_t$, the critical exponents exhibit crossover behavior.  

At the tricritical point, the critical exponent values of $\delta^\prime$, $\eta$, $z$, and $\delta$ are obtained for each value of $\sigma$ in the range $[0.1,3]$ in steps of $\Delta \sigma=0.1$, as shown in Fig.~\ref{fig:fig9}(b). The obtained critical values are listed in Table~\ref{tab:tab3}. Each critical exponent value exhibits crossover behavior across $\sigma_{c1}$ and $\sigma_{c2}$. The value of $\sigma_{c1}$ is determined to be $4/3$ in two dimensions, because $d_c=1.5\sigma_{c1}$. For $\sigma < \sigma_{c1}$, mean-field behavior occurs, whereas for $\sigma > \sigma_{c1}$, a significant low-dimensional fluctuation effect appears. 
The upper bound $\sigma_{c2}$, across which the universality class changes from the two-dimensional LTDP class to the two-dimensional STDP class, was determined using the hyperscaling relation~\eqref{eq:hyperscaling}.
We remark that whereas in the regions $\sigma < \sigma_{c1}$ and $\sigma > \sigma_{c2}$, the exponents are constant regardless of $\sigma$, in the interval $[\sigma_{c1},\,\sigma_{c2}]$, the critical exponents vary constantly as a function of $\sigma$, which is a prototypical pattern that appears in the long-range CP model.

Indeed, we find numerically that the critical exponent values for $\sigma$ between $ [\sigma_{c1}=4/3,\sigma_{c2}\approx 2.2]$ vary depending on $\sigma$, as listed in Table~\ref{tab:tab3}. For instance, for $\sigma=2.0$, we obtain the critical exponents directly by measuring the slopes as $\eta=-0.129\pm 0.010$, $\delta^\prime=1.073\pm 0.010$, $z=1.840\pm 0.015$, and $\delta=0.212\pm 0.010$, as shown in Fig.~\ref{fig:fig10}. We also obtain the critical exponents using the FSS method. We plot $N_a t^{-\eta}$ versus $t N^{-\bar{z}}$ for different system sizes $N$ in Fig.~\ref{fig:fig11}(a), the rescaled quantity $P(t)t^{\delta'}$ versus $tN^{-\bar{z}}$ in Fig.~\ref{fig:fig11}(b).
The exponent $\nu_{\|}$ is obtained from the scaling plot of $N_a(t)t^{-\eta}$ versus $t^{1/\nu_{\|}}(\kappa_t-\kappa)$ for different values of $\kappa$ in Fig.~\ref{fig:fig12}(a). The data points for different $\kappa$ values collapse well onto the curve for $\nu_{\|}=1.07\pm 0.005$. In Fig.~\ref{fig:fig12}(b), the crossover exponent $\phi$ is obtained from the scaling plot of $\rho_{a,s}(\omega_t-\omega)^{-\beta/\phi}$ versus $(\kappa-\kappa_c)(\omega_t-\omega)^{-1/\phi}$ for different values of $\omega$. The data points for different values of $\omega$ also collapse well onto a curve for $\phi=0.52\pm 0.02$. The critical exponent values for other $
\sigma$ values are listed in Table~\ref{tab:tab3}.


Using the field theory approach, the upper critical dimension was determined to be $d_c=1.5\sigma$~\cite{lastWork}. Thus, when $\sigma < 4/3$, $d_c$ is smaller than $d=2$.
In this case, when we perform dimensional analysis, we need to use $d_c$ rather than $d=2$. For instance, for hyperscaling analysis, we need to use $\bar{\nu}=d_c\nu$, i.e., $\bar{\nu}=(1.5\sigma)\nu$ for $\sigma < 4/3$ and $\bar{\nu}=2\nu$ for $\sigma > 4/3$ in two dimensions. 
To confirm this scaling theory, {for $\sigma=2.0>4/3$}, we obtain the dynamic exponent $z$ by directly measuring the local slope of the plot of $R^2(t)$ versus $t$ in Fig.~\ref{fig:fig10}(c) and the exponent $\bar z$ from the scaling plots in Figs.~\ref{fig:fig11}(a) and (b).
{For $\sigma<4/3$, we also obtain the dynamic exponent $z$ in Fig.~\ref{fig:fig13}(a) and the exponent $\bar z$ from the scaling plots in Fig.~\ref{fig:fig13}(b).}  
{Thus,} we confirm that {$z/\bar{z}$} is close to $2$ for $\sigma=2$ and $1.204$ for $\sigma=0.8$.
The hysteresis of the first-order transition for $\omega > \omega_t$ is shown in Fig.~\ref{fig:fig14}.

\subsection{LTCP model in one dimension}
\label{sec:5-4}
\begin{figure}[ht!]
\includegraphics[width=.85\linewidth]{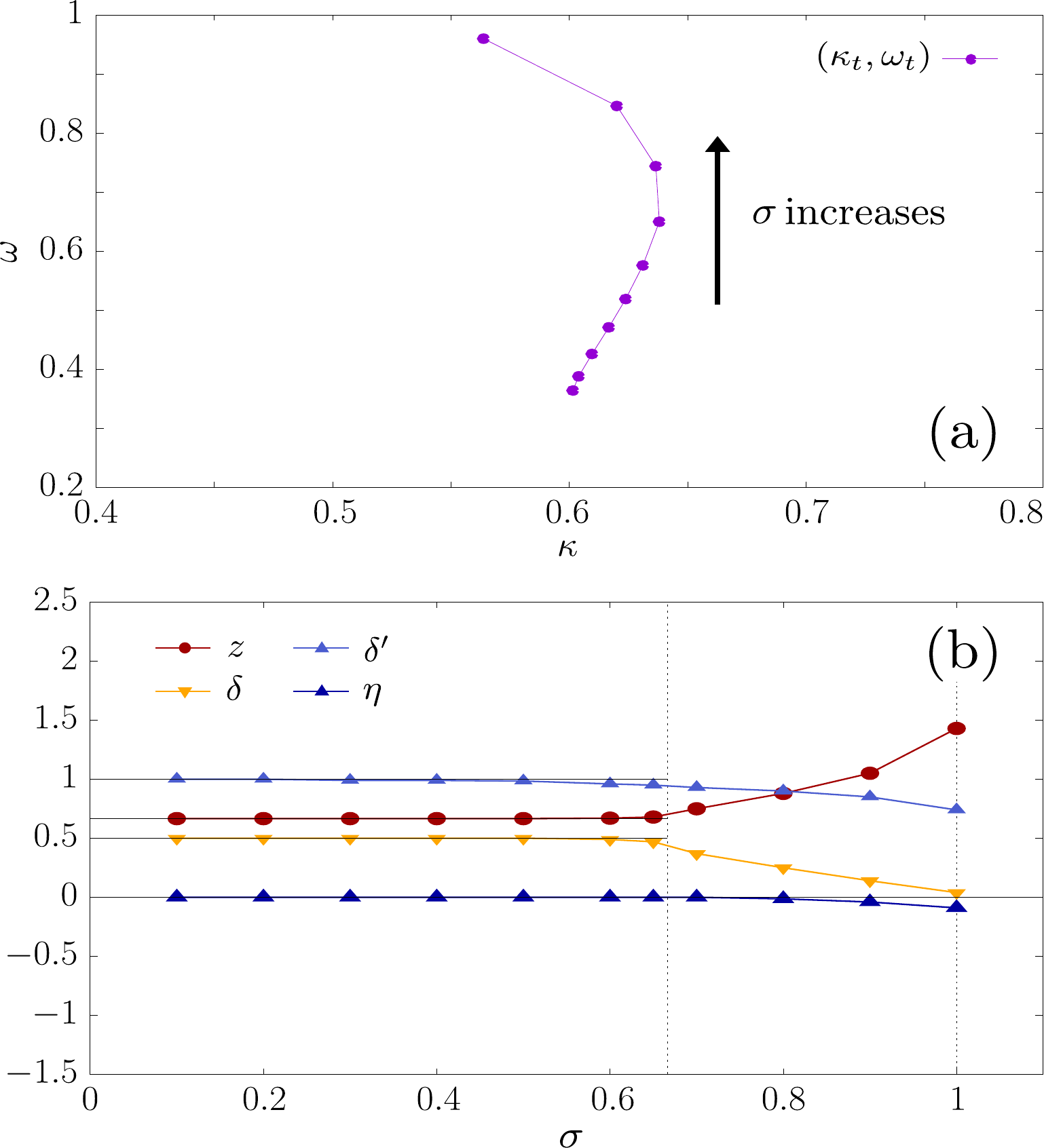}
\caption{For the LTCP model in one dimension, (a) plot of the tricritical points in ($\kappa,\omega$) space for different $\sigma$ values in $[0.1,1.0]$. (b) Plots of the critical exponent values $z$, $\delta$, $\delta^\prime$, and $\eta$ as a function of $\sigma$ at the tricritical point. $\sigma_{c1}$ and $\sigma_{c2}$ are indicated by vertical dotted lines. 
\label{fig:fig15}}
\end{figure}

\begin{figure*}[ht!]
\includegraphics[width=1.75\columnwidth]{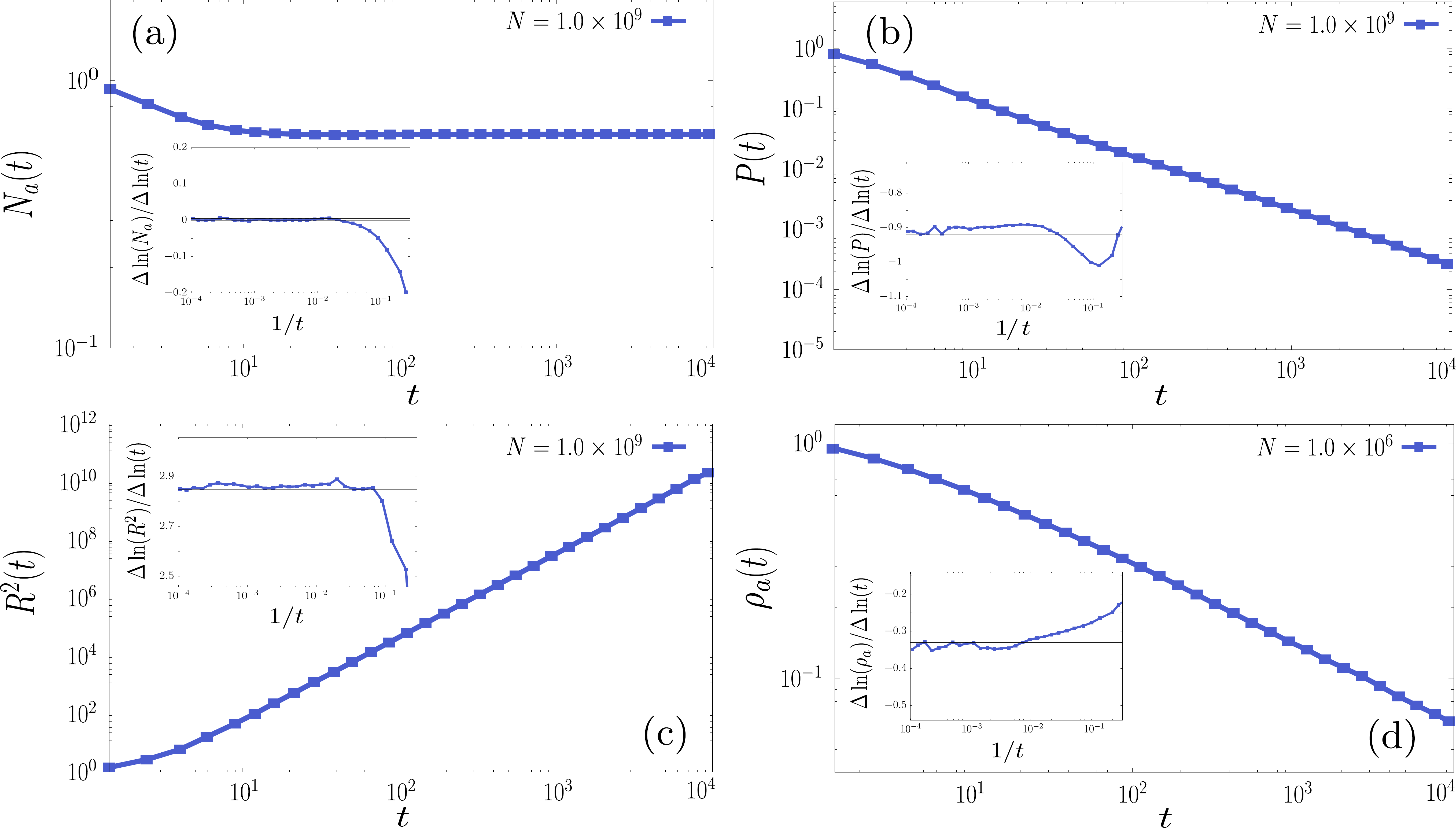}
\caption{For the LTDP model with $\sigma=0.7$ in one dimension at the tricritical point, plots of (a) $N_a(t)$, (b) $P(t)$, (c) $R^2(t)$, and (d) $\rho_a(t)$ versus $t$. We estimate the exponent values to be $\eta=0.000\pm 0.005$, $\delta^\prime=0.912\pm 0.01$, $z=0.701\pm 0.01$, and $\delta=0.34\pm 0.01$, respectively. Insets: local slopes of each quantity versus $1/t$. 
\label{fig:fig16}}
\end{figure*}

\begin{figure}[ht!]
\includegraphics[width=0.85\columnwidth]{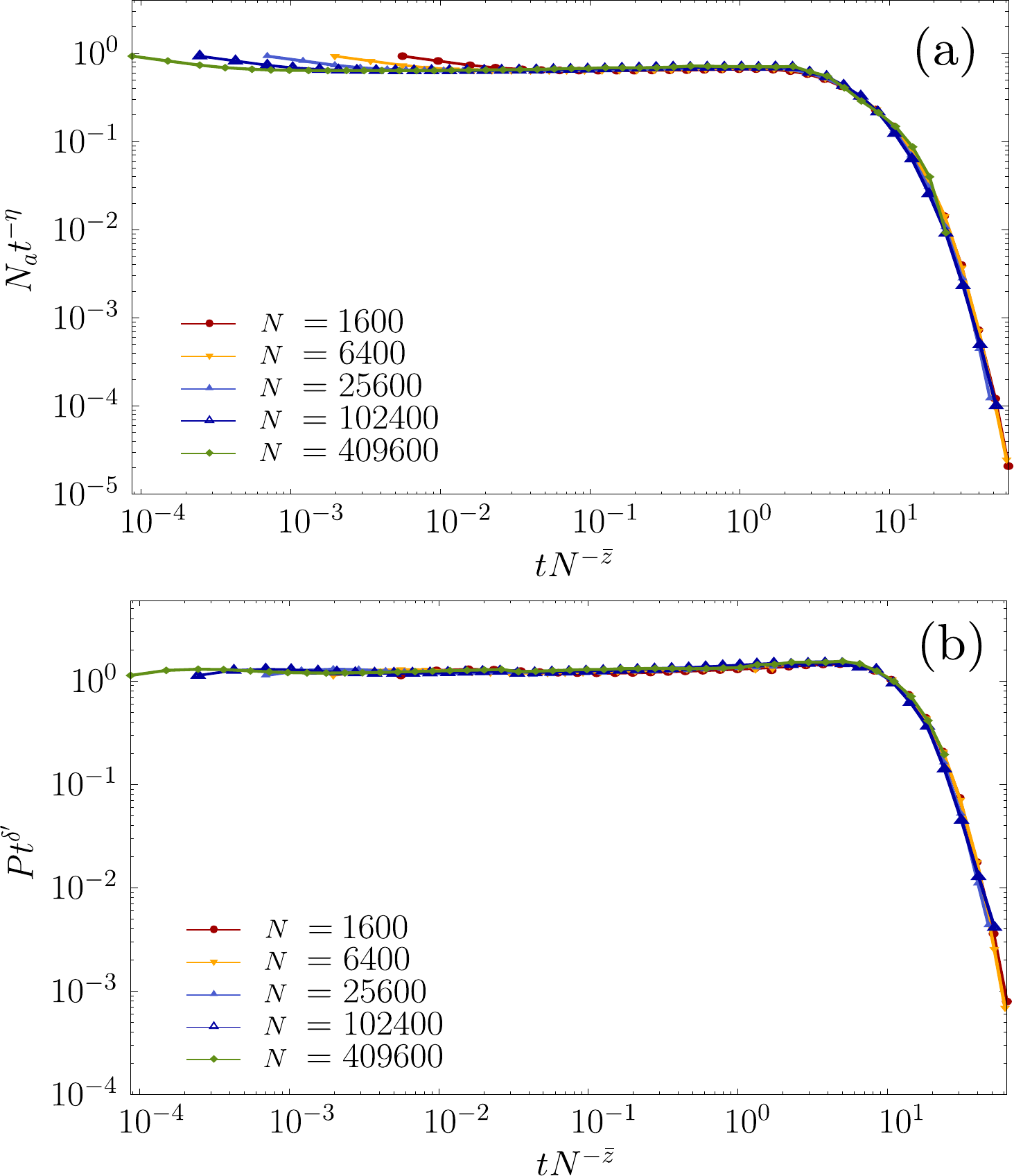}
\caption{For the LTDP model with $\sigma=0.7$ in one dimension, (a) scaling plot of $N_at^{-\eta}$ versus $tN^{-\bar{z}}$ for $\eta=0.000$ and $\bar{z}=0.701$. (b) Scaling plot of $Pt^{\delta^\prime}$ versus $tN^{-\bar{z}}$ for $\delta^\prime=0.912$ and $\bar{z}=0.701$. 
\label{fig:fig17}}
\end{figure}

\begin{figure}[ht!]
\includegraphics[width=0.85\columnwidth]{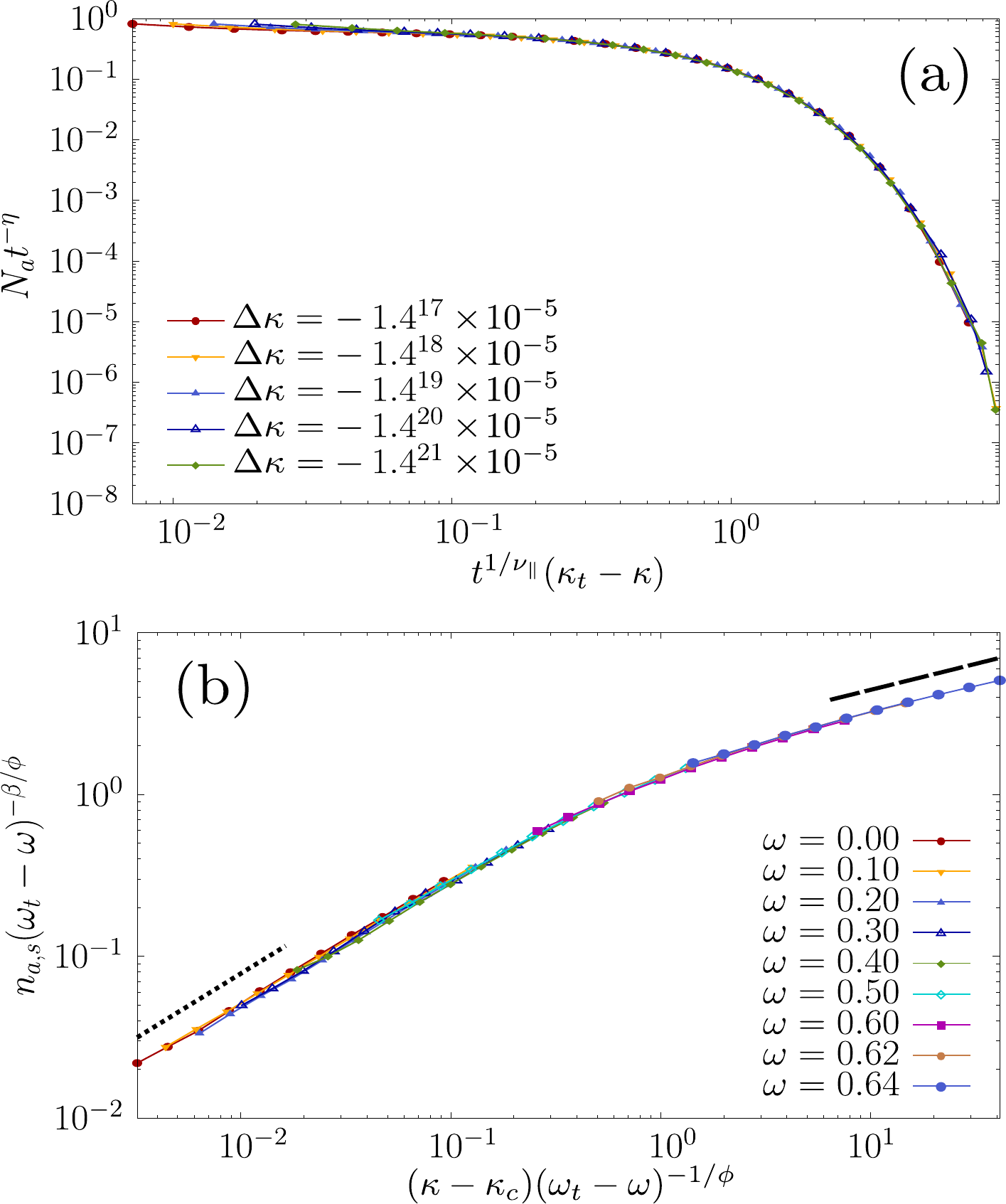}
\caption{For the LTCP model with $\sigma=0.7$ in one dimension, (a) scaling plot of $N_at^{-\eta}$ versus $t^{1/\nu_{\|}}(\kappa_t-\kappa)$ for different values of $\kappa$. Data points collapse well onto a single curve for $\kappa_t=0.637508$, $\eta=0.00$, and $\nu_{\|}=1.05$. (b) Scaling plot of $\rho_{a,s}(\omega_t-\omega)^{-\beta/\phi}$ versus $(\kappa-\kappa_c)(\omega_t-\omega)^{-1/\phi}$ for different values of $\omega$. Dotted (Dashed) line is a guideline with slope $\beta_{\rm LDP}=0.800$ ($\beta_{t}=0.321$). Data points collapse well onto a single curve with $\phi=0.52$. 
\label{fig:fig18}}
\end{figure}

\begin{figure}[ht!]
\includegraphics[width=0.85\columnwidth]{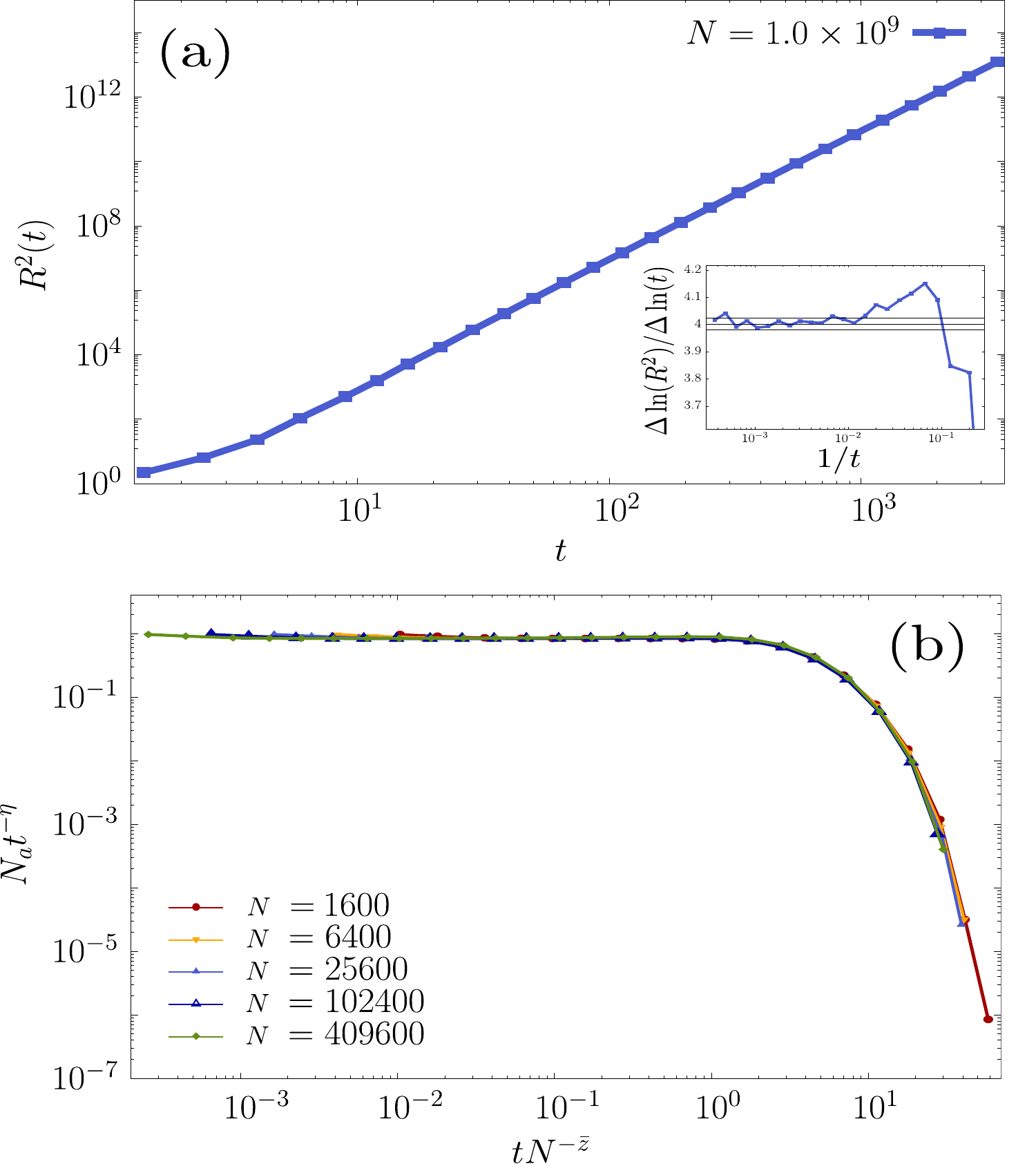}
\caption{For the LTCP model with $\sigma=0.5$ in one dimension, (a) plot of $R^2(t)$ versus $t$. We obtain the exponent $2/z=4.004\pm 0.010$. Inset represents local slopes at each $t$ as a function of $1/t$. (b) Scaling plot of $N_a t^{-\eta}$ versus $tN^{-\bar{z}}$ for $\eta=0$ and $\bar{z}=0.666$.} 
\label{fig:fig19}
\end{figure}

\begin{figure}[ht!]
\includegraphics[width=0.85\columnwidth]{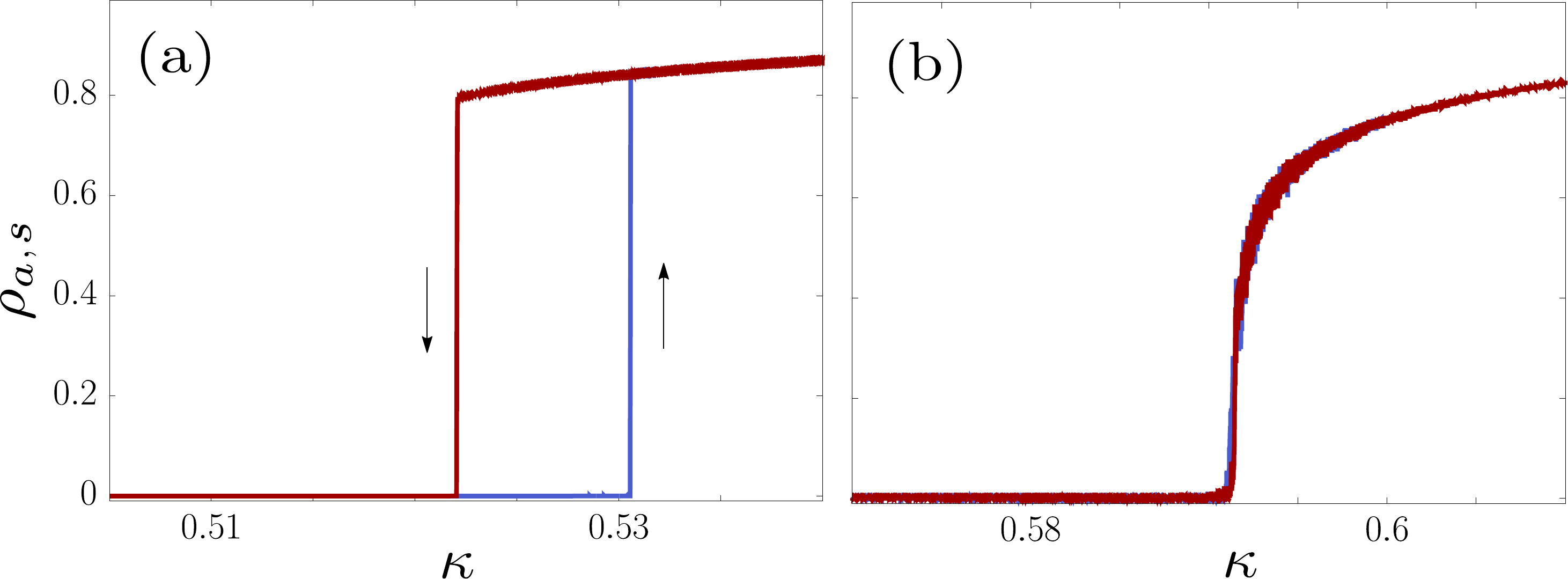}
\caption{Plot of $\rho_{a,s}$ versus $\kappa$ for the LTCP model in one dimension. (a) With $\sigma=0.7$ at $\omega=0.9 > \omega_t$, a hysteresis curve is obtained. (b) With $\sigma=1.0$ at $\omega=0.97$, a hysteresis curve does not occur. The system size is taken as $N=10^6$.}
\label{fig:fig20}
\end{figure}

We perform numerical simulations in one dimension, in which the exponent $\sigma$ was taken from the interval $[0.1,\,1.1]$ in steps of $\Delta \sigma=0.1$. For each value of $\sigma$, we determine a critical point $(\kappa_c, \omega_c)$ and a tricritical point $(\kappa_t,\,\omega_t)$ using the same method as in the previous subsections. As in the phase diagram of the $m$-TCP model, a second-order (first-order) transition occurs for $\omega < \omega_t(\sigma)$ ($\omega > \omega_t(\sigma)$).

Next, we determine an interval $[\sigma_{c1}, \sigma_{c2}]$ within which the dynamics of the LTCP model becomes nontrivial. We first determine that $\sigma_{c1}=2/3$ using the criterion $d=1.5\sigma_{c1}$ at $d=1$. For $\sigma < \sigma_{c1}$, the mean-field solution is valid, and the upper critical dimension is determined as $d_c=(3/2)\sigma$. For $\sigma > \sigma_{c1}$, the upper critical dimension is larger than the dimension $d=1$. Accordingly, when we use $\bar{\nu}=d\nu$, we need to take $\bar{\nu}=(1.5\sigma)\nu$ for $\sigma < 2/3$ and $\bar{\nu}=\nu$ for $\sigma > 2/3$ in one dimension. {We confirm this property by measuring the ratio $\bar z/z$ for different $\sigma$ values smaller than and larger than $\sigma_{c1}$. }
{To confirm this scaling theory, for $\sigma=0.7>2/3$, we obtain the dynamic exponent $z$ by directly measuring the local slope in the plot of $R^2(t)$ versus $t$ in Fig.~\ref{fig:fig16}(c) and the exponent $\bar z$ from the scaling plots in Figs.~\ref{fig:fig17}(a) and (b).
For $\sigma<2/3$, we also obtain the dynamic exponent $z$ in Fig.~\ref{fig:fig19}(a) and the exponent $\bar z$ from the scaling plots in Fig.~\ref{fig:fig19}(b).}  
{Thus, we confirm that $z/\bar{z}$ is close to $1$ for $\sigma=0.7$ and $0.749$ for $\sigma=0.5$.}

To determine $\sigma_{c2}$, we recall the previous result that for a short-range DP-type CP model, a first-order transition does not occur in one dimension~\cite{hinrichsen_first}. Thus, an STDP class does not appear in the region $\sigma > \sigma_{c2}$~\cite{windus_1D}.  
On the basis of this background, we need to determine the $\sigma$ range in which the LTCP universality class exists in one dimension. Thus, we need to check whether a tricritical point exists in the interval $\sigma_{c1}<\sigma<\sigma_{c2}$.

The numerical simulation results show that a tricritical point still exists in the region $\sigma>\sigma_{c1}$, but it disappears near $\sigma\simeq 1.0$, as a discontinuous transition does not occur (Fig.~\ref{fig:fig20}). Thus, we take $\sigma_{c2} \simeq 1.0$. The tricritical points for given $\sigma$ values less than $\sigma_{c2}$ are determined and shown in Fig.~\ref{fig:fig15}(a). As $\sigma$ approaches $1.0$ in the phase diagram, $\omega_t$ also approaches 1.0 [Fig.~\ref{fig:fig15}(a)]. By contrast, when $\omega=1$, the LTDP dynamics is frozen because the absorbing state is reached immediately after the dynamics starts from an initial configuration in which either all the sites are fully active or only one site is active (see Table~\ref{tab:tab1}). Thus, the dynamics near $\sigma\approx 1.0$ is so sensitive that precise numerical measurement of the critical exponents is almost impossible.

At the tricritical point, the critical exponent values of $\delta^\prime$, $\eta$, $z$, and $\delta$ are obtained for each value of $\sigma$ in the range $[0.1,1.0]$ in steps of $\Delta \sigma=0.1$, as shown in Fig.~\ref{fig:fig15}(b). The obtained critical values are listed in Table~\ref{tab:tab4}. For $\sigma > \sigma_{c2}$, all the transition lines belongs to the DP class. We remark that whereas in the region $\sigma < \sigma_{c1}$, the exponents are constant regardless of $\sigma$, they vary constantly as a function of $\sigma$ in the interval $[\sigma_{c1},\,\sigma_{c2}]$, which is a prototypical pattern that appears in the long-range CP model. Indeed, we find numerically that the critical exponent values for $\sigma$ between $ [\sigma_{c1}=2/3,\sigma_{c2}\approx 1.0]$ vary depending on $\sigma$, as shown in Table~\ref{tab:tab4}. 
For instance, for $\sigma=0.7$, we obtain the critical exponents directly by slope measurements as $\eta=0.000\pm 0.005$, $\delta^\prime=0.912\pm 0.01$, $z=0.701\pm 0.01$, and $\delta=0.34\pm 0.01$, as shown in Fig.~\ref{fig:fig16}. We also obtain the critical exponents using the FSS method. We plot $N_a t^{-\eta}$ versus $t N^{-\bar{z}}$ for different system sizes $N$ in Fig.~\ref{fig:fig17}(a), the rescaled quantity $P(t)t^{\delta'}$ versus $tN^{-\bar{z}}$ in Fig.~\ref{fig:fig17}(b).
The exponent $\nu_{\|}$ is obtained from the scaling plot of $N_a(t)t^{-\eta}$ versus $t^{1/\nu_{\|}}(\kappa_t-\kappa)$ for different values of $\kappa$ in Fig.~\ref{fig:fig18}(a). The data points for different $\kappa$ values collapse well onto a curve for $\nu_{\|}=1.05\pm 0.005$. In Fig.~\ref{fig:fig18}(b), the crossover exponent $\phi$ is obtained from the scaling plot of $\rho_{a,s}(\omega_t-\omega)^{-\beta/\phi}$ versus $(\kappa-\kappa_c)(\omega_t-\omega)^{-1/\phi}$ for different values of $\omega$. The data points for different values of $\omega$ also collapse well onto a curve for $\phi=0.52\pm 0.02$. The critical exponent values for other $\sigma$ values are listed in Table~\ref{tab:tab4}. The hysteresis of the first-order transition for $\omega > \omega_t$ is shown in Fig.~\ref{fig:fig20}.

\section{Conclusion and Discussion}
\label{sec:6}
\label{sec:conclusion}

\begin{figure}[ht!]
\includegraphics[width=0.85\columnwidth]{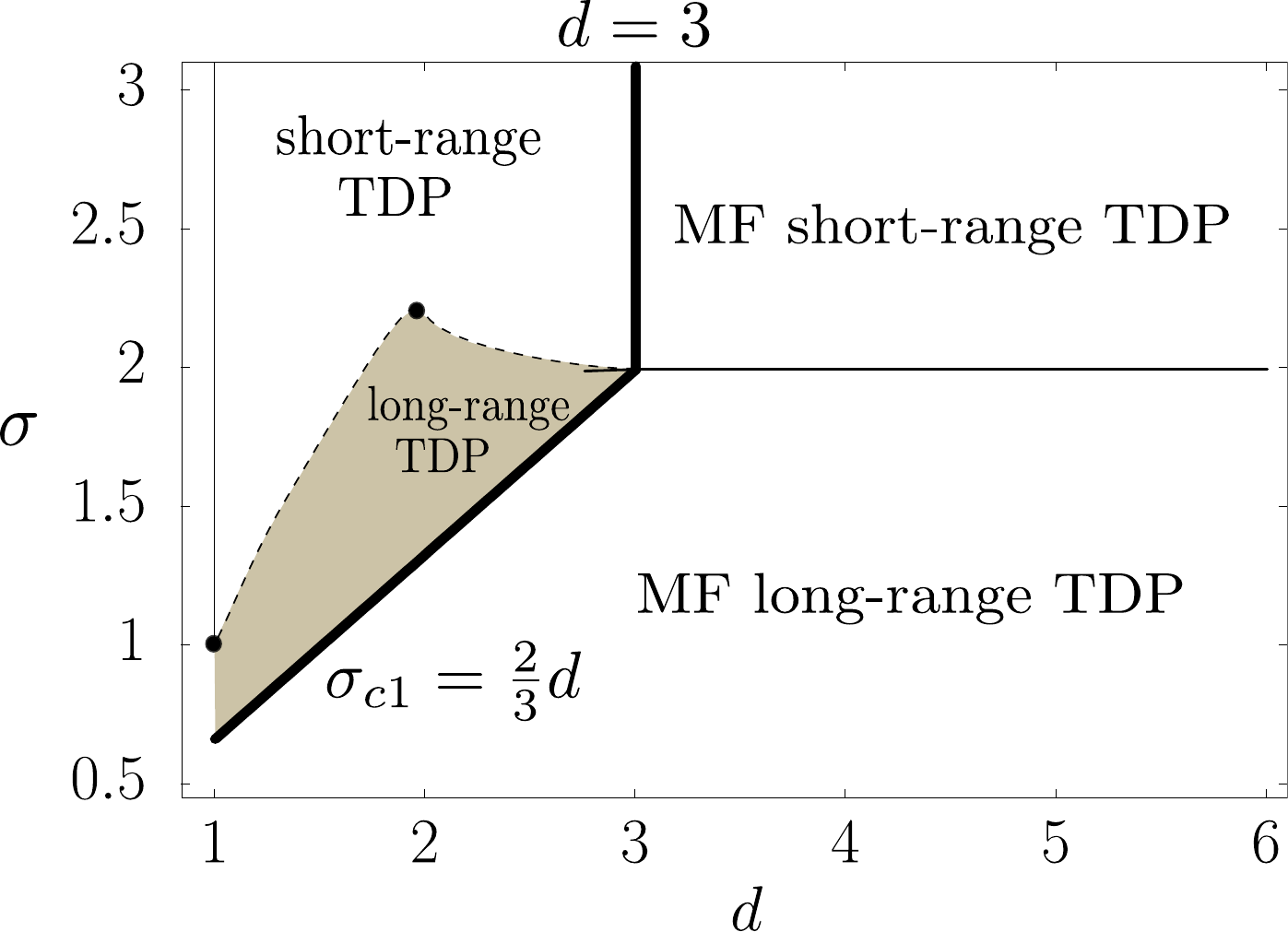}
\caption{Diagram of universality classes of the LTCP model in the parameter space ($d, \sigma$). Mean-field solution is valid beyond the upper critical dimension line (bold line), min$(3,\,1.5\sigma)$. The slope of the dashed line near $(d,\, \sigma)=(3,\,2)$ is $0.0304$, according to Eq.~\eqref{eq:hyperscaling}, and is indicated by a short solid line. The dot at $(2,\,2.2)$ indicates the $\sigma_{c2}$ value obtained from numerical simulations of the LTCP model in two dimensions. The dot at $(1,\,1.0)$ was numerically estimated for $d=1$ and indicates $\sigma_{c2}$. The dashed curves connecting these three points separate the STDP region from the LTDP region. Along the thin solid line above the point (1,1) in one dimension, a tricritical point is absent, so this thin line is excluded from the STDP region.}
	\label{fig:fig21}
\end{figure}

In this paper, we investigated the critical behavior of the LTCP model, i.e., the TCP with long-range interaction in the form of $1/r^{d+\sigma}$ at a tricritical point, in one and two dimensions. First, we determined the domain of the LTDP universality class in the parameter space ($d,\,\sigma$), as shown in Fig.~\ref{fig:fig21}. The domain is surrounded by the domains of the low-dimensional STDP class, the mean-field STDP class, and the mean-field LTDP class. These four domains meet at the point $(d,\sigma)=(3,\,2)$. Below the upper critical dimension $d_c=3$, the domain of the LTDP class is sandwiched between those of the low-dimensional STDP and the mean-field LTDP class, denoted as the shaded area bounded by $\sigma_{c1}(d) < \sigma < \sigma_{c2}(d)$ for each $d$. Analytically, $\sigma_{c1}(d)$ was determined using the formula $\sigma_{c1}=(2/3)d$, which was derived by dimensional analysis in the mean-field limit. $\sigma_{c2}(d)$ was determined using $d+z(1-\delta-\delta^\prime)$, where $z$, $\delta$, and $\delta^\prime$ are the exponents of the STDP class. Near the point ($3,\,2$), using the $\epsilon$ expansion of the RG approach for the STDP class~\cite{ohtsuki1,janssen_TDP}, we obtained $\sigma=2-0.0304\epsilon+\mathcal{O}(\epsilon^2)$ in $d=3-\epsilon$ spatial dimensions. Therefore, we obtained the tangent of the phase boundary between the STDP and LTDP domains as $\Delta\sigma/\Delta d \approx 0.0304$ near the point $(3,\,2)$ in Fig.~\ref{fig:fig21}.   
 
Second, we numerically determined the critical exponent values at the tricritical point for $d=1$ and $d=2$. The numerical results showed that although the critical exponents are independent of the control parameter $\sigma$ for $\sigma < \sigma_{c1}$ and $\sigma > \sigma_{c2}$, they vary continuously with $\sigma$ between [$\sigma_{c1}, \sigma_{c2}$]. The numerical values of the critical exponents are listed in Tables~\ref{tab:tab3} and \ref{tab:tab4} (Appendix~\ref{appendixC}). 

We unexpectedly obtained the following noteworthy behavior. First, in two dimensions, the numerically obtained value of $\sigma_{c2}$ was not consistent with the theoretical value based on the STDP class; rather, it was close to the value obtained using the DP class. Second, in one dimension, the boundary $\sigma_{c2}$ could not be determined from the STDP class, because the first-order transition does not occur in one dimension for the ordinary CP model. Thus, we determined $\sigma_{c2}$ only numerically. 

The LTCP model in one dimension is particularly notable. For each given $\sigma$ in the range [$0.1,1.0]$ in steps of $\Delta \sigma=0.1$, as shown in Fig.~\ref{fig:fig15}(a), there exists a tricritical point $(\kappa_t,\omega_t)$. This figure shows that $\omega_t$ increases with increasing $\sigma$. We also found that the gap in the discontinuous transition near the tricritical point at $\sigma=0.7$ (Fig.~\ref{fig:fig20} (a)) is supposed to be decreasing as increasing $\sigma$ and eventually the gap diminishes at a characteristic value of $\sigma$,  denoted as $\sigma_{c2}$ and estimated to be $\sigma_{c2}\approx 1.0$ (Fig.~\ref{fig:fig20} (b)). This implies that $\omega_t$ approaches $\omega_t\to 1$, the upper bound of the pair-branching probability $\omega$. 
{In the semi-classical approach, the quantum coherence effect was regarded as the classical effect of a pair-branching process with the control parameter $\omega$.}
{Thus when $\omega=1$, the LTCP model most highly reflects the quantum coherence effect. On the other hand, the previous studies~\cite{carollo, gillman, roscher} of QCP in one dimension revealed that the tricritical point does not occur. This previous result seems to be associated with the current result that the tricritical point disappears beyond $\omega=1$ as $\sigma > 1$. This is also consistent with another previous result that the STCP model, corresponding to the limit $\sigma\to \infty$ of the LTCP model, does not exhibit any discontinuous transition~\cite{hinrichsen_first} in one dimension owing to strong fluctuation effect.}   

{In this respect, although a discontinuous transition was not observed in short-range QCP in one dimension~\cite{carollo, gillman}, we guess that it could occur in long-range QCP in one dimension. In this case, a tricritical point and a new emerging behavior could be observed. An experiment of Rydberg atoms exciting to $d$-state is a potential candidate. Due to dipole interactions, long-range interaction is intrinsically generated.}

In summary, we obtained the diagram of universality classes based on the analytical and numerical results in Tables~\ref{tab:tab3} and \ref{tab:tab4} (Fig.~\ref{fig:fig21}). The local slope at $d=3$ was determined by inserting the results of $\epsilon$ expansion for the STDP class~\cite{ohtsuki1,janssen_TDP} into Eq.~\eqref{eq:hyperscaling}. The values of $\sigma_{c2}$ in one and two dimensions were numerically obtained. In Ref.~\cite{kahng}, the discrepancy between the simulated and field-theoretical $\sigma_{c2}$ values is reported for other models such as the Ising~\cite{blanchard,picco} and percolation models~\cite{grassberger_SIR_2d}. Here, although we obtained the Monte Carlo simulation results, the $\epsilon$ expansion of the LTCP model is still  missing. Thus, further studies are needed from the perspective of RG theory.

\begin{acknowledgments}
This research was supported by the NRF, Grant No.~NRF-2014R1A3A2069005 (BK).
\end{acknowledgments}

\begin{appendix}

\begin{widetext}
\section{Calculation of propagator}\label{appendixB}
Let us evaluate the propagator loop integral to determine why the fractional Laplacian is not renormalized. This can be done using the propagator loop integral in Ref.~\cite{ohtsuki1}. The only difference between the ordinary TDP and the LTDP lies in the Green function, which changes slightly from $G(\bm{k},{\omega})=(Dk^2-i{\omega}\tau+u_2)^{-1}$ to $G(\bm{k},{\omega})=(D_{\sigma}k^{\sigma}-i{\omega}\tau+u_2)^{-1}$.
Because the cubic terms remain the same, the relevant diagrams do not change, as shown in Ref.~\cite{ohtsuki1}. $I_0(k,\omega)$ is given by
\begin{align}\label{eq:originalI0}
I_0(k,\omega) &=
\begin{gathered}
\includegraphics[width=0.05\linewidth]{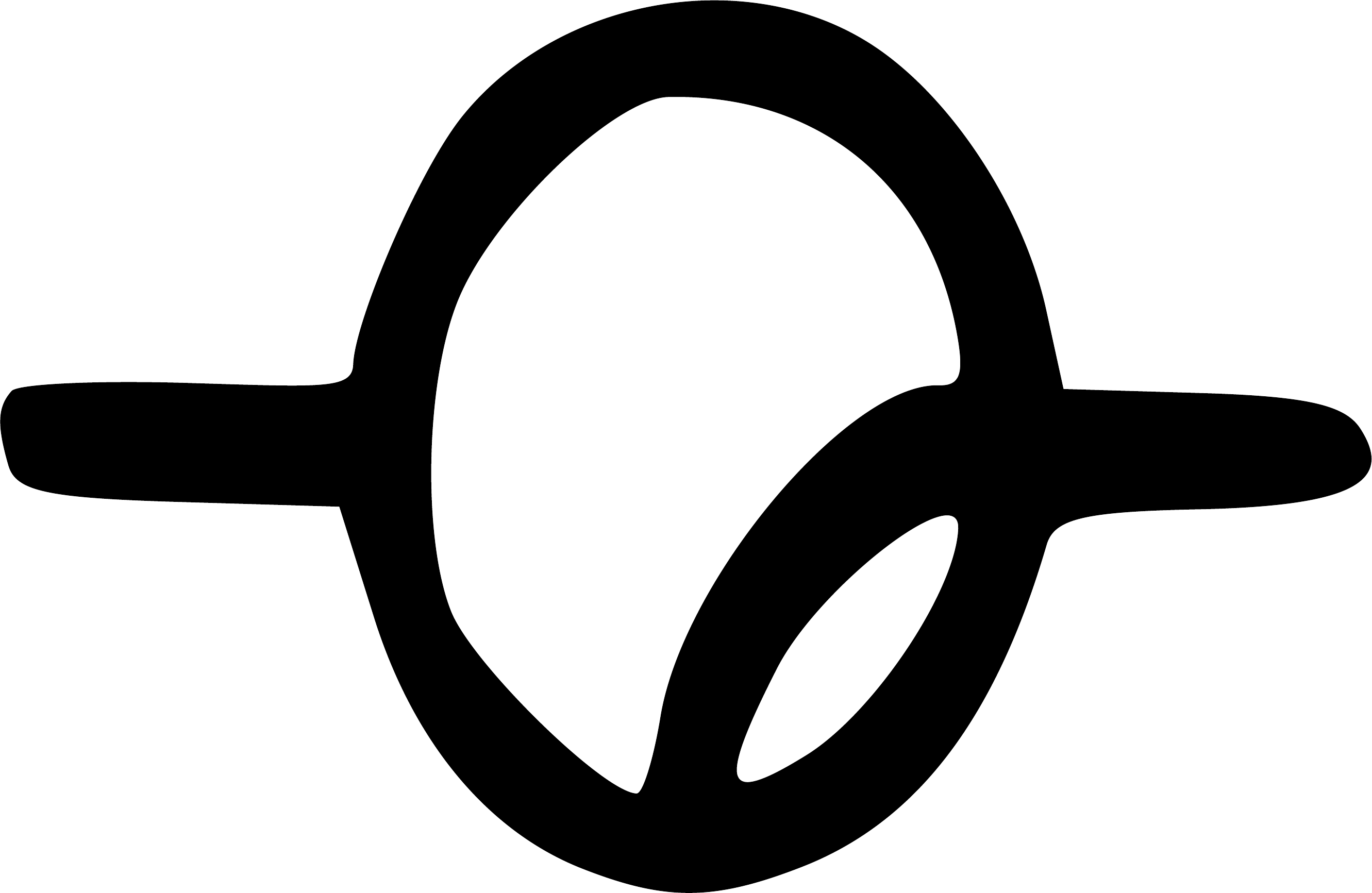}
\end{gathered}
\nonumber\\
&=\int \frac{d^dq_1}{(2\pi)^{d}}\int \frac{d^dq_2}{(2\pi)^{d}}
\int \frac{d\omega_1}{(2\pi)}\int \frac{d\omega_2}{(2\pi)}
G(-\bm{k}-\bm{q}_1,-\omega-\omega_1)G(\bm{q}_1,\omega_1)G(\bm{q}_1+\bm{q}_2,\omega_1+\omega_2)G(-\bm{q}_2,-\omega_2)\nonumber\\
&=
\int \frac{d^dq_1}{(2\pi)^{d}}\int \frac{d^dq_2}{(2\pi)^{d}}
\int \frac{d\omega_1}{(2\pi)}
\frac{1}{D_{\sigma}|\bm{k}+\bm{q}_1|^{\sigma}+i({\omega}+\omega_1)\tau+u_2}
\frac{1}{D_{\sigma}q_1^{\sigma}-i{\omega_1}\tau+u_2}
\frac{1}{D_{\sigma}q^{\sigma}+D_{\sigma}|\bm{q}_1+\bm{q}_2|^{\sigma}-i{\omega_1}\tau+2u_2}\nonumber\\
&=\int \frac{d^dq_1}{(2\pi)^{d}}\int \frac{d^dq_2}{(2\pi)^{d}}\frac{1}{(i\omega+D_{\sigma}q_1^{\sigma}+D_{\sigma}|\bm{q}_1+\bm{k}|^{\sigma}+2u_2)(i\omega+D_{\sigma}|\bm{q}_1+\bm{q}_2|^{\sigma}+D_{\sigma}q^{\sigma}+D_{\sigma}|\bm{q}_1+\bm{k}|^{\sigma}+3u_2)}\,,
\end{align}
where we set $\tau=1$ and $D=1$ without loss of generality. 
After $\omega_1$ and $\omega_2$ in Eq.~\eqref{eq:originalI0} are integrated out using the Cauchy integral, $I_0$ is given by
\begin{align}\label{eq:I0}
I_0(k,\omega) 
&=
\int \frac{d^dq_1}{(2\pi)^{d}}\int \frac{d^dq_2}{(2\pi)^{d}}\frac{1}{(i\omega+q_1^{\sigma}+|\bm{q}_1+\bm{k}|^{\sigma}+2u_2)(i\omega+|\bm{q}_1+\bm{q}_2|^{\sigma}+q_2^{\sigma}+|\bm{q}_1+\bm{k}|^{\sigma}+3u_2)}\nonumber\\
&=\int \frac{d^dq_1}{(2\pi)^{d}}\int \frac{d^dq_2}{(2\pi)^{d}}
\frac{1}{i\omega+q_1^{\sigma}+|\bm{q}_1+\bm{k}|^{\sigma}+2u_2}
\int^{i\infty}_{-i\infty}dz_1
\Gamma(1+z_1)\Gamma(-z_1)\frac{\Big(i\omega+q_2^{\sigma}+|\bm{q}_1+\bm{k}|^{\sigma}+3u_2\Big)^{z_1}}{|\bm{q}_1+\bm{q}_2|^{\sigma(1+z_1)}}
\nonumber\\
&=
\int \frac{d^dq_1}{(2\pi)^{d}}\int \frac{d^dq_2}{(2\pi)^{d}}
\frac{1}{i\omega+q_1^{\sigma}+|\bm{q}_1+\bm{k}|^{\sigma}+2u_2}
\int^{i\infty}_{-i\infty}dz_1
\int^{i\infty}_{-i\infty}dz_2
\Gamma(1+z_1)\Gamma(-z_1)\Gamma(z_2-z_1)\Gamma(-z_2)
\frac{\Big(i\omega+|\bm{q}_1+\bm{k}|^{\sigma}+3u_2\Big)^{z_2}}{|\bm{q}_1+\bm{q}_2|^{\sigma(1+z_1)}q_2^{\sigma(z_2-z_1)}}\,,
\end{align}
where we used the Mellin--Barnes representation $$\frac{1}{(X+Y)^{\lambda}}=\int_{-i\infty}^{i\infty}dz \frac{Y^z}{X^{\lambda+z}}\frac{\Gamma(\lambda+z)\Gamma(-z)}{\Gamma(\lambda)}\,.$$ 
Now, the integral over $q_2$ in Eq.~\eqref{eq:I0} becomes
\begin{align}\label{formula}
\int\frac{d^dq_2}{(2\pi)^d}\frac{1}{q_2^a|\bm{q}_1+\bm{q}_2|^b}
&=\frac{q_1^{d-(a+b)}\Gamma(\frac{a+b-d}{2})\Gamma(\frac{d-a}{2})\Gamma(\frac{d-b}{2})}{(4\pi)^{d/2}\Gamma(\frac{a}{2})\Gamma(\frac{b}{2})\Gamma(d-\frac{a+b}{2})}\,.
\end{align}
After Eq.~\eqref{formula} is inserted into Eq.~\eqref{eq:I0}, $I_0$ is given by
\begin{align}\label{eq:I0_1}
I_0=\int \frac{d^dq_1}{(16\pi^3)^{d/2}}
\int^{i\infty}_{-i\infty}dz_1
\int^{i\infty}_{-i\infty}dz_2
&\Gamma(1+z_1)\Gamma(-z_1)\Gamma(z_2-z_1)\Gamma(-z_2)\,\times\nonumber\\
&\frac{\Gamma(\frac{\sigma}{2}(z_2+1)-\frac{d}{2})
\Gamma(\frac{d}{2}-\frac{\sigma}{2}(z_1+1))
\Gamma(\frac{d}{2}-\frac{\sigma}{2}(z_2-z_1))
}{
\Gamma(\frac{\sigma}{2}(1+z_1))
\Gamma(\frac{\sigma}{2}(z_2-z_1))
\Gamma(d-\frac{\sigma}{2}(1+z_2))
}
\frac{q_1^{d-\sigma(z_2+1)}\Big(i\omega+|\bm{q}_1+\bm{k}|^{\sigma}+3u_2\Big)^{z_2}}{i\omega+q_1^{\sigma}+|\bm{q}_1+\bm{k}|^{\sigma}+2u_2}\,.
\end{align}
Then, let us expand the last term in Eq.~\eqref{eq:I0_1} with respect to $k$ and $\omega$.
\begin{align}\label{eq:I0_2}
&\int \frac{d^dq_1}{(16\pi^3)^{d/2}}
\frac{q_1^{d-\sigma(z_2+1)}\Big(i\omega+|\bm{q}_1+\bm{k}|^{\sigma}+3u_2\Big)^{z_2}}{i\omega+q_1^{\sigma}+|\bm{q}_1+\bm{k}|^{\sigma}+2u_2} \nonumber\\
&=\int \frac{d^dq_1}{(16\pi^3)^{d/2}}q_1^{d-\sigma(z_2+1)}\frac{\Big(q_1^{\sigma}+3u_2+i\omega+\sigma q_1^{\sigma-1}k\cos(\theta) +\frac{\sigma}{2}k^2q_1^{\sigma-2}+\frac{\sigma}{4}(\frac{\sigma}{2}-1)q_1^{\sigma-2}k^2\cos^2(\theta)+\mathcal{O}(k^3)\Big)^{z_2}}{2q_1^{\sigma}+2u_2+i\omega+\sigma q_1^{\sigma-1}k\cos(\theta) +\frac{\sigma}{2}k^2q_1^{\sigma-2}+\frac{\sigma}{4}(\frac{\sigma}{2}-1)q_1^{\sigma-2}k^2\cos^2(\theta)+\mathcal{O}(k^3)}\nonumber\\
&=\int \frac{d^dq_1}{(16\pi^3)^{d/2}}\frac{q_1^{d-\sigma(z_2+1)}(3u_2+q_1^{\sigma})^{z_2}}{2q_1^{\sigma}+2u_2}\Big(1+\frac{i\omega+\sigma q_1^{\sigma-1}k\cos(\theta) +\frac{\sigma}{2}k^2q_1^{\sigma-2}+\frac{\sigma}{4}(\frac{\sigma}{2}-1)q_1^{\sigma-2}k^2\cos^2(\theta)+\mathcal{O}(k^3)}{3u_2+q_1^{\sigma}}\Big)^{z_2}\,\times\nonumber\\
&\Big(1+\frac{i\omega+\sigma q_1^{\sigma-1}k\cos(\theta) +\frac{\sigma}{2}k^2q_1^{\sigma-2}+\frac{\sigma}{4}(\frac{\sigma}{2}-1)q_1^{\sigma-2}k^2\cos^2(\theta)+\mathcal{O}(k^3)}{2q_1^{\sigma}+2u_2}\Big)^{-1}\nonumber\\
&=\int \frac{d^dq_1}{(16\pi^3)^{d/2}}\frac{q_1^{d-\sigma(z_2+1)}(3u_2+q_1^{\sigma})^{z_2}}{2q_1^{\sigma}+2u_2}\Big[ 1+i\omega\Big( \frac{z_2}{3u_2+q_1^{\sigma}}-\frac{1}{2u_2+2q_1^{\sigma}}  \Big)+k\Big( \frac{z_2\sigma q_1^{\sigma-1}\cos(\theta)}{3u_2+q_1^{\sigma}}-\frac{\sigma q_1^{\sigma-1}\cos(\theta)}{2u_2+2q_1^{\sigma}} \Big)\nonumber\\
&+k^2\Big(   \frac{z_2\Big(\frac{\sigma}{2} q_1^{\sigma-2}+\frac{\sigma}{4}(\frac{\sigma}{2}-1)\cos^2(\theta)\Big)}{3u_2+q_1^{\sigma}}-\frac{\frac{\sigma}{2} q_1^{\sigma-2}+\frac{\sigma}{4}(\frac{\sigma}{2}-1)\cos^2(\theta)}{2u_2+2q_1^{\sigma}}+\frac{z_2(z_2-1)\sigma^2 q_1^{2\sigma-2}\cos^2(\theta)}{2(3u_2+q_1^{\sigma})^2}\nonumber\\
&+\frac{\sigma^2 q_1^{2\sigma-2}\cos^2(\theta)}{(2u_2+2q_1^{\sigma})^2}+\frac{z_2\sigma^2 q_1^{2\sigma-2}\cos^2(\theta)}{(3u_2+q_1^{\sigma})(2u_2+2q_1^{\sigma})} \Big)+\mathcal{O}(k^3,\omega^2,k\omega)\Big]\,.
\end{align}
We used the following relation, because $k$ is very small in the long-wavelength limit.
\begin{align}
(\bm{q}_1+\bm{k})^{\sigma}&=((\bm{q}_1+\bm{k})^2)^{\sigma/2}\nonumber\\
&=q_1^{\sigma}+\sigma q_1^{\sigma-2}\bm{q}_1\cdot\bm{k} +\frac{\sigma}{2}k^2q_1^{\sigma-2}+\frac{\sigma}{4}(\frac{\sigma}{2}-1)q_1^{\sigma-4}(2\bm{q}_1\cdot\bm{k})^2+\mathcal{O}(k^3)\nonumber\\
&=q_1^{\sigma}+\sigma q_1^{\sigma-1}k\cos(\theta) +\frac{\sigma}{2}k^2q_1^{\sigma-2}
+\sigma(\frac{\sigma}{2}-1)q_1^{\sigma-2}k^2\cos^2(\theta)+\mathcal{O}(k^3)\,,
\end{align}
where $\theta$ is the angle between $\bm{q}_1$ and $\bm{k}$. To evaluate Eq.~\eqref{eq:I0_2}, it is often helpful to use the formulas
\begin{align}
\int \frac{d^dq_1}{(2\pi)^d}f(q_1)
&=\frac{S_{d-1}}{(2\pi)^{d}}\int_0^{\infty}dq_1\int_0^{\pi}d\theta f(q_1)\sin^{d-2}(\theta) 
=\frac{S_{d}}{(2\pi)^{d}}\int_0^{\infty}dq_1 f(q_1)\,,\nonumber\\
\int \frac{d^dq_1}{(2\pi)^d}f(q_1) \cos(\theta)
&=\frac{S_{d-1}}{(2\pi)^{d}}\int_0^{\infty}dq_1\int_0^{\pi}d\theta f(q_1)\sin^{d-2}(\theta) \cos(\theta)=0\,,\nonumber\\
\int \frac{d^dq_1}{(2\pi)^d}f(q_1) \cos^2(\theta)
&=\frac{S_{d-1}}{(2\pi)^{d}}\int_0^{\infty}dq_1\int_0^{\pi}d\theta f(q_1)\sin^{d-2}(\theta) \cos^2(\theta)
=\frac{S_{d}}{d(2\pi)^{d}}\int_0^{\infty}dq_1 f(q_1)\,,
\end{align}
where the surface area is defined as $S_d=\frac{2\pi^{d/2}}{\Gamma(d/2)}$. Then, Eq.~\eqref{eq:I0_1} is given as follows:
\begin{align}\label{eq:I0_3}
I_0(k,\omega)&=
\int^{i\infty}_{-i\infty}dz_1
\int^{i\infty}_{-i\infty}dz_2
\Gamma(1+z_1)\Gamma(-z_1)\Gamma(z_2-z_1)\Gamma(-z_2)\,\times\nonumber\\
&\frac{\Gamma(\frac{\sigma}{2}(z_2+1)-\frac{d}{2})
\Gamma(\frac{d}{2}-\frac{\sigma}{2}(z_1+1))
\Gamma(\frac{d}{2}-\frac{\sigma}{2}(z_2-z_1))
}{
\Gamma(\frac{\sigma}{2}(1+z_1))
\Gamma(\frac{\sigma}{2}(z_2-z_1))
\Gamma(d-\frac{\sigma}{2}(1+z_2))
}
S_d\int_0^{\infty} \frac{dq_1}{(16\pi^3)^{d/2}}
\frac{q_1^{d-\sigma(z_2+1)}(3u_2+q_1^{\sigma})^{z_2}}{2q_1^{\sigma}+2u_2} \nonumber\\
&\Big[1+\frac{i\omega}{D_{\sigma}}\Big(  \frac{z_2}{3u_2+q_1^{\sigma}}-\frac{1}{2u_2+2q_1^{\sigma}}  \Big)
+k^2\Big(   \frac{z_2\Big(\frac{\sigma}{2} q_1^{\sigma-2}+\frac{\sigma}{4d}(\frac{\sigma}{2}-1)\Big)}{3u_2+q_1^{\sigma}}-\frac{\frac{\sigma}{2} q_1^{\sigma-2}+\frac{\sigma}{4d}(\frac{\sigma}{2}-1)}{2u_2+2q_1^{\sigma}}\nonumber\\
&
+\frac{z_2(z_2-1)\sigma^2 q_1^{2\sigma-2}}{2d(3u_2+q_1^{\sigma})^2}+\frac{\sigma^2 q_1^{2\sigma-2}}{d(2u_2+2q_1^{\sigma})^2}
+\frac{z_2\sigma^2 q_1^{2\sigma-2}}{d(3u_2+q_1^{\sigma})(2u_2+2q_1^{\sigma})} \Big)\Big]
+\mathcal{O}(k^3,\omega^2,k\omega)
 \nonumber\\
&= N_0+N_{\omega} \omega + N_{k^2} k^2+\mathcal{O}(k^3,\omega^2, k\omega)
\,,
\end{align}
where $N_0$, $N_{\omega}$, and $N_{k^2}$ are coefficients.
Finally, because the derivative of Eq.~\eqref{eq:I0_3} with respect to $k^{\sigma}$ vanishes, the coefficient $D_{\sigma}$ is not renormalized up to {the} first order in the $\epsilon$ expansion around the upper critical dimension. Although we showed that it is valid for up to $\mathcal{O}(\epsilon)$, it is commonly believed that nonlocal terms of the dynamic action are not renormalized at all~\cite{notRenormalized,notRenormalized2,notRenormalized3}.
\end{widetext}

\section{Tables of numerical estimates}
\label{appendixC}
\begin{table*}[h!]
\begin{center}
\caption{Critical exponents for the LTDP model in two dimensions.}
\setlength{\tabcolsep}{4pt}
{\renewcommand{\arraystretch}{1.5}
\begin{tabular}{cccccccccc}
    \hline
    \hline
    $\sigma$ & $(\kappa_t,\,\omega_t)$ & $\delta$ & $\delta^\prime$ & $\bar{z}\equiv z/d$ & $\nu_{\|}$ & $\eta$ & $\phi$ & Universality \\
   \hline
      $0.2$ & $(0.609401,\,0.378)$         & $0.500\pm 0.005$& $1.00\pm 0.02$&   $0.666\pm 0.01$ & $1.00\pm 0.01$& $0.000 \pm 0.005$    &$0.50\pm 0.01$ &\\
   $0.4$ & $(0.612271,\,0.400)$         & $0.500\pm 0.005$& $1.00\pm 0.02$&   $0.666\pm 0.01$ & $1.00\pm 0.01$& $0.000 \pm 0.005$    &$0.50\pm 0.01$ &\\
   $0.6$ & $(0.616819,\,0.424)$         & $0.500\pm 0.005$& $1.00\pm 0.02$&   $0.666\pm 0.01$ & $1.00\pm 0.01$& $0.000 \pm 0.005$    &$0.50\pm 0.01$ &Mean-field long-range TDP\\
   $0.8$ & $(0.622538,\,0.450)$         & $0.500\pm 0.005$& $1.00\pm 0.02$&   $0.666\pm 0.01$ & $1.00\pm 0.01$& $0.000 \pm 0.005$    &$0.50\pm 0.01$ &\\
   $1.0$ & $(0.628244,\,0.475)$         & $0.500\pm 0.01$&  $0.99\pm 0.02$&   $0.666\pm 0.01$ & $1.00\pm 0.01$& $0.000 \pm 0.005$    &$0.50\pm 0.01$ &\\
   $1.2$ & $(0.635410,\,0.506)$         & $0.485\pm 0.01$&  $1.00\pm 0.02$&   $0.678\pm 0.01$ & $1.02\pm 0.01$& $-0.010\pm 0.005$   &$0.50\pm 0.01$ &\\\hline
   $1.4$ & $(0.643351,\,0.543)$         & $0.397\pm 0.01$&  $1.01\pm 0.02$&   $0.725\pm 0.01$ & $1.03\pm 0.01$& $-0.022\pm 0.005$   &$0.50\pm 0.01$ &  \\
   $1.5$ & $(0.647071,\,0.562)$         & $0.345\pm 0.01$&  $1.013\pm 0.02$&   $0.758\pm 0.01$ & $1.03\pm 0.01$& $-0.029\pm 0.005$   &$0.51\pm 0.01$ &  \\
   $1.6$ & $(0.650679,\,0.582)$         & $0.309\pm 0.01$&  $1.021\pm 0.02$&   $0.784\pm 0.01$ & $1.04\pm 0.01$& $-0.032\pm 0.005$   &$0.51\pm 0.01$ & \\
   $1.7$ & $(0.653822,\,0.601)$         & $0.281\pm 0.01$&  $1.031\pm 0.02$&   $0.819\pm 0.01$ & $1.04\pm 0.01$& $-0.052\pm 0.005$   &$0.51\pm 0.01$ & \\
   $1.8$ & $(0.656771,\,0.621)$         & $0.253\pm 0.01$&  $1.041\pm 0.02$&   $0.849\pm 0.01$ & $1.05\pm 0.01$& $-0.071\pm 0.01$   &$0.51\pm 0.02$ &Long-range TDP\\
   $1.9$ & $(0.659371,\,0.641)$         & $0.223\pm 0.01$&  $1.050\pm 0.01$&   $0.881\pm 0.01$ & $1.06\pm 0.01$& $-0.091\pm 0.01$   &$0.52\pm 0.02$ & \\
   $2.0$ & $(0.661659,\,0.662)$         & $0.212\pm 0.01$&  $1.073\pm 0.01$&   $0.922\pm 0.01$ & $1.07\pm 0.01$& $-0.129\pm 0.01$   &$0.52\pm 0.02$ &  \\
   $2.1$ & $(0.663511,\,0.683)$         & $0.184\pm 0.01$&  $1.100\pm 0.01$&   $0.961\pm 0.01$ & $1.08\pm 0.01$& $-0.180\pm 0.01$   &$0.52\pm 0.02$ & \\
   $2.2$ & $(0.664880,\,0.703)$         & $0.172\pm 0.01$&  $1.150\pm 0.01$&   $0.992\pm 0.01$ & $1.09\pm 0.01$& $-0.211\pm 0.01$   &$0.52\pm 0.02$ & \\\hline
   $2.4$ & $(0.666269,\,0.742)$         & $0.123\pm 0.01$&  $1.211\pm 0.01$&   $1.045\pm 0.01$ & $1.12\pm 0.01$& $-0.288\pm 0.01$   &$0.52\pm 0.02$ & \\
   $2.6$ & $(0.666487,\,0.769)$         & $0.105\pm 0.01$&  $1.22\pm 0.01$&    $1.055\pm 0.01$ & $1.14\pm 0.01$& $-0.334\pm 0.01$   &$0.52\pm 0.02$ & Short-range TDP \\
   $2.8$ & $(0.666001,\,0.792)$         & $0.098\pm 0.01$&  $1.22\pm 0.01$&    $1.055\pm 0.01$ & $1.15\pm 0.01$& $-0.353\pm 0.01$   &$0.52\pm 0.02$ &  \\
   $3.0$ & $(0.665661,\,0.806)$         & $0.089\pm 0.01$&  $1.22\pm 0.01$&    $1.055\pm 0.01$ & $1.15\pm 0.01$& $-0.353\pm 0.01$   &$0.52\pm 0.02$ & \\
   $\infty$ & $(0.6606466,\,0.879)$      & $0.09\pm 0.01$& $1.22\pm 0.008$&    $1.055\pm 0.005$ & $1.15\pm 0.005$& $-0.35 \pm 0.008$   &$0.52\pm 0.02$ &  \\
  \hline
    \hline
\end{tabular}}
\label{tab:tab3}
\end{center}
\end{table*}

\begin{table*}[h!]
\begin{center}
\caption{Critical exponents for the LTDP model in one dimension.
}
\setlength{\tabcolsep}{4pt}
{\renewcommand{\arraystretch}{1.5}
\begin{tabular}{cccccccccc}
    \hline
    \hline
    $\sigma$ & $(\kappa_t,\omega_t)$ & $\delta$ & $\delta^\prime$ & $\bar{z}\equiv z/d$ & $\nu_{\|}$ & $\eta$ & $\phi$ & Universality \\
   \hline
  $0.2$ & $(0.603919,\,0.388)$ &$0.50\pm 0.005$& $1.000\pm 0.01$& $0.666\pm 0.005$& $1.00\pm 0.01 $& $0.00\pm 0.005$  & $0.50\pm 0.02$&\\
  $0.4$ & $(0.616681,\,0.471)$ &$0.50\pm 0.005$& $0.99\pm 0.01$& $0.666\pm 0.005$ & $1.00\pm 0.01$& $0.00\pm 0.005$   & $0.50\pm 0.02$ &Mean-field long-range TDP\\
  $0.6$ & $(0.631031,\,0.576)$ &$0.49\pm 0.01$& $0.96\pm 0.01$& $0.670\pm 0.005$ & $1.01\pm 0.01$& $0.00\pm 0.005$   & $0.50\pm 0.02$&\\ \hline
  $0.7$ & $(0.637510,\,0.654)$ &$0.34\pm 0.01$& $0.91\pm 0.01$& $0.701\pm 0.01$ & $1.05\pm 0.01$& $0.00\pm 0.005$   & $0.52\pm 0.02$ &\\
  $0.8$ & $(0.637551,\,0.744)$ &$0.25\pm 0.01$& $0.88\pm 0.01$& $0.878\pm 0.01$ & $1.09\pm 0.01$& $-0.013\pm 0.01$ & $0.54\pm 0.02$ & \\
  $0.9$ & $(0.622539,\,0.846)$ &$0.14\pm 0.01$& $0.83\pm 0.01$& $1.05 \pm 0.01$ & $1.18\pm 0.01$& $-0.04\pm 0.01$  & $0.56\pm 0.02$& Long-range TDP \\
  $1.0$ & $(0.556705,\,0.960)$ &$0.04\pm 0.01$& $0.76\pm 0.01$& $1.43 \pm 0.01$ & $1.34\pm 0.01$& $-0.09\pm 0.01$  & $0.58\pm 0.02$ & \\\hline
  $1.05$ &  \multicolumn{8}{c}{ Tricritical point does not exist.}& \\
    \hline
    \hline
\end{tabular}}
\label{tab:tab4}
\end{center}
\end{table*}

\end{appendix}

\end{document}